\documentclass[11pt,a4paper]{article}
\usepackage{epsfig}
\usepackage[T1]{fontenc}    
\usepackage{graphics}
\usepackage{graphicx}
\usepackage{pstricks,pst-coil,pst-fill,pst-plot}
\usepackage[fleqn]{amsmath}    
\usepackage{amssymb}    
\usepackage{amsfonts}   
\usepackage{verbatim}   
\usepackage{mathrsfs}   
\usepackage{dsfont}
\usepackage{euscript}
\usepackage{yfonts}
\usepackage{enumerate}     
\usepackage{amsthm}         
\usepackage{txfonts}
\usepackage{marvosym}
\usepackage{stmaryrd}
\usepackage{vmargin}        
\usepackage{wasysym}		
\usepackage{navigator}

\setmarginsrb{1.8cm}{2cm}{1.8cm}{2cm}{1cm}{1cm}{1cm}{1.6cm}
 \makeatletter
 \@addtoreset{equation}{section}
 \makeatother

\providecommand{\bysame}{\leavevmode\hbox to3em{\hrulefill}\thinspace}
\providecommand{\MR}{\relax\ifhmode\unskip\space\fi MR }

\providecommand{\href}[2]{#2}

\let\tend=\rightarrow

\long\def\symbolfootnote[#1]#2{\begingroup%
\def\thefootnote{\fnsymbol{footnote}}\footnote[#1]{#2}\endgroup}

\newtheorem{theorem}{Theorem}[section]
\newtheorem{prop}[theorem]{Proposition}
\newtheorem*{theorem*}{Theorem}

\newtheorem{lemme}[theorem]{Lemma}

\def\Proof{\medskip\noindent {\it Proof --- \ }}

\def\qed{\hfill\rule{2mm}{2mm}}

\newcommand\beq{\begin{equation}}
\newcommand\enq{\end{equation}}
\newcommand\bem{\begin{multline}}
\newcommand\enm{\end{multline}}

\def\beqa{\begin{eqnarray}}
\def\eeqa{\end{eqnarray}}
\def\ba{\begin{array}}
\def\ea{\end{array}}
\def\det{\operatorname{det}}

\newcommand{\f}[2]{{\ensuremath{%
    \mathchoice%
    {\dfrac{#1}{#2}}
    {\dfrac{#1}{#2}}
    {\frac{#1}{#2}}
    {\frac{#1}{#2}}
}}}
\newcommand{\tf}[2]{\ensuremath{#1/#2}}
\newcommand{\pa}[1]{\ensuremath{\left(#1\right)}}

\def\be{\beta}

\def\Ga{\Gamma}

\def\de{\delta}

\def\De{\Delta}
\def\eps{\epsilon}

\def\la{\lambda}
\def\La{\Lambda}

\def\sg{\sigma}

\def\th{\theta}

\def\om{\omega}

\newcommand{\mc}[1]{\ensuremath{\mathcal{#1}}}
\newcommand{\mf}[1]{\ensuremath{\mathfrak{#1}}}
\newcommand{\msc}[1]{\ensuremath{\mathscr{#1}}}

\newcommand{\bs}[1]{\ensuremath{\boldsymbol{#1}}}

\DeclareFontFamily{OT1}{pzc}{}
\DeclareFontShape{OT1}{pzc}{m}{it}{<-> s * [1.10] pzcmi7t}{}
\DeclareMathAlphabet{\mathpzc}{OT1}{pzc}{m}{it}

\def \i{ \mathrm i}

\newcommand{\ov}[1]{\ensuremath{\overline{#1}}}
\newcommand{\wt}[1]{\ensuremath{\widetilde{#1}}}
\newcommand{\wh}[1]{\ensuremath{\widehat{#1}}}

\newcommand{\Int}[2]{\ensuremath{\int\limits_{#1}^{#2}}}
\newcommand{\Oint}[2]{\ensuremath{\oint\limits_{#1}^{#2}}}

\newcommand{\sul}[2]{\ensuremath{\sum\limits_{#1}^{#2}}}
\newcommand{\pl}[2]{\ensuremath{\prod\limits_{#1}^{#2}}}

\newcommand{\R}{\ensuremath{\mathbb{R}}}
\newcommand{\Cx}{\ensuremath{\mathbb{C}}}

\newcommand{\Dp}[1]{\ensuremath{\partial_{#1}}}

\newcommand{\limit}[2]{\ensuremath{\underset{#1 \tend #2}{\longrightarrow} }}

\newcommand{\s}[1]{\ensuremath{\sinh\pa{#1}}}

\newcommand{\ex}[1]{\ensuremath{\e{e}^{#1}}}

\newcommand{\op}[1]{ \boldsymbol{ \texttt{#1} } }

\newcommand{\dd}{\mathrm{d}}
\newcommand{\e}[1]{\ensuremath{\mathrm{#1}}}

\newcommand{\intfo}[2]{\ensuremath{ [  #1 \,; #2 [ }}

\newcommand{\widesim}[2][1.5]{
  \mathrel{\underset{#2}{\scalebox{#1}[1]{$\sim$}}}
}

\begin{document}

\begin{center}
\begin{LARGE}
{\bf Multi-point correlation functions in the boundary XXZ chain\\ at finite temperature}
\end{LARGE}

\vspace{1cm}

\vspace{4mm}
{\large Karol K. Kozlowski \footnote{e-mail: karol.kozlowski@ens-lyon.fr}}%
\\[1ex]
Univ Lyon, ENS de Lyon, Univ Claude Bernard Lyon 1, CNRS, Laboratoire de Physique, F-69342 Lyon, France \\[2.5ex]

\vspace{4mm}
{\large V\'{e}ronique Terras \footnote{e-mail: veronique.terras@universite-paris-saclay.fr}}%
\\[1ex]
Université Paris-Saclay, CNRS, LPTMS, 91405, Orsay, France \\[2.5ex]

\par 

\vspace{4mm}

\today
\vspace{40pt}

\end{center}


\centerline{\bf Abstract} \vspace{4mm}
\begin{center}

\parbox{12cm}{\small 
We consider multi-point correlation functions in the open XXZ chain with longitudinal boundary fields and in a uniform external magnetic field. We show that, at finite temperature, these correlation functions can be written in the quantum transfer matrix framework as sums over thermal form factors. More precisely, and quite remarkably, each term of the sum is given by a simple product of usual matrix elements of the quantum transfer matrix multiplied by a unique factor containing the whole information about the boundary fields.
As an example, we provide a detailed expression for the longitudinal spin one-point functions at distance $m$ from the boundary. This work thus solves the long-standing problem of setting up form
factor expansions in integrable models subject to open boundary conditions.}

\end{center}

\vspace{40pt}

\tableofcontents

\section{Introduction}

The calculation of the correlation functions in interacting integrable models is a long-standing problem. 
The first attempts in this direction goes back to the works of Takahashi \cite{TakahashiSpinSpinSecondNeighborXXX}: the latter developed ingenious roundabout arguments leading to closed expressions for some of the next-to-neighbouring correlation functions in the XXX chain, the one of the nearest neighbouring spin operator following trivially from Hulten's result \cite{HultenGSandEnergyForXXX} for the model's ground state energy. 
It however turned out to be very difficult to go beyond these special cases in the framework of the coordinate Bethe Ansatz. In fact, genuine progress could only be made after the formulation of the algebraic Bethe Ansatz \cite{FaddeevSklyaninTakhtajanSineGordonFieldModel}. 
Then, two structurally different approaches to the problem emerged. The first approach goes back to the works of the Kyoto school \cite{JimboMikiMiwaNakayashikiElementaryBlocksXXZperiodicDelta>1,JimboMiwaFormFactorsInMassiveXXZ,JimboMiwaElementaryBlocksXXZperiodicMassless} for the XXZ chain in the infinite volume limit: there, the $U_{q}(\wh{\mf{sl}}_2)$ symmetry of the infinite
XXZ chain was used so as to set up a vertex operator approach and systems of q-KZ equations, leading to the computation of the form factor densities of local operators in the massive regime of the XXZ chain \cite{JimboMiwaFormFactorsInMassiveXXZ}
along with the zero temperature reduced density matrix  \cite{JimboMikiMiwaNakayashikiElementaryBlocksXXZperiodicDelta>1,JimboMiwaElementaryBlocksXXZperiodicMassless} 
in all regimes of the XXZ chain.  
The \textit{per se} algebraic Bethe Ansatz approach to the calculation of correlation functions in integrable models in finite volume was pioneered in the works \cite{IzerginKorepinQISMApproachToCorrFns2SiteModel,IzerginKorepinQISMApproachToCorrNextDiscussion}, see also \cite{BogoliubiovIzerginKorepinBookCorrFctAndABA}. 
However, in this early stage, the obtained representations for the correlators suffered from an important combinatorial intricacy. This problem was later overcome 
thanks to the obtention of a convenient determinant representation for the scalar product of off-shell/on-shell Bethe vectors \cite{SlavnovScalarProductsXXZ}  along with
the resolution of the so-called quantum inverse scattering problem \cite{KitanineMailletTerrasFormfactorsperiodicXXZ}. Eventually, the algebraic Bethe Ansatz  approach gave rise to various types of series of multiple integral representations for the 
correlation functions, both in finite volume and in the thermodynamic limit
\cite{KozKitMailSlaTer6VertexRMatrixMasterEquation,KitanineMailletSlavnovTerrasOriginalSeries,KitanineMailletSlavnovTerrasDynamicalCorrelationFunctions,KitanineMailletSlavnovTerrasMasterEquation}.
In particular, it reproduced the integral representations 
for the zero temperature reduced density matrix of the XXZ infinite chain previously obtained in the framework of the aforementioned first approach, 
and extended them to the case of finite overall magnetic fields \cite{KitanineMailletTerrasElementaryBlocksPeriodicXXZ}.
>From these representations it was eventually possible to extract, without any \textit{ad hoc} hypothesis or handling, the long-distance asymptotic behaviour of two-point functions in the massless regime of the chain \cite{KozKitMailSlaTerXXZsgZsgZAsymptotics}.
Later on, it turned out that the determinant representations  for the form factors of local operators in the XXZ chain obtained in \cite{KitanineMailletTerrasFormfactorsperiodicXXZ}
could be analysed  in the large-volume, thermodynamic, limit \cite{KozKitMailSlaTerEffectiveFormFactorsForXXZ,KozKitMailSlaTerThermoLimPartHoleFormFactorsForXXZ}, which then made it possible  to  efficiently describe, 
by means of form factor expansions, the dynamical two-point correlation functions, as well as to  grasp their 
critical  behaviours \cite{KozKitMailSlaTerRestrictedSums,KozKitMailSlaTerRestrictedSumsEdgeAndLongTime,KozMasslessFFSeriesXXZ,KozLongDistanceLargeTimeXXZ,KozSingularitiesSpectralFctsXXZ}. 
Moreover, by using the quantum transfer matrix formalism, it was also possible to generalise these approaches to the case of the correlation functions at finite temperature \cite{GohmannKlumperSeelFinieTemperatureCorrelationFunctionsXXZ}.
In particular, the thermal form factor expansions \cite{KozDugaveGohmannThermaxFormFactorsXXZ,KozDugaveGohmannThermaxFormFactorsXXZOffTransverseFunctions}  turned out to be most efficient for studying the critical regime.  

All the results mentioned so far pertained to the XXZ spin $1/2$ chain or to the one-dimensional Bose gaz subject to periodic boundary conditions. 
The two mentioned approaches to the computation of correlation functions could also be adapted, at least to some extent, to deal with other kinds of integrable boundary conditions. 
It particular, it has been possible to provide  closed expressions for the zero temperature reduced density matrix \cite{JimboKedemKonnoMiwaXXZChainWithaBoundaryElemBlcks,KozKitMailNicSlaTerElemntaryblocksopenXXZ} and the two-point correlation functions \cite{KozKitMAilNicSlaTerResummationsOpenXXZ}
of the XXZ chain subject to diagonal boundary fields. Unfortunately,  these representations did not turn out to be efficient enough for the analysis of the long-distance asymptotic behaviour of the correlation functions. 
 Also, the lack of translational invariance hindered the implementation of the form factor approach to criticality, an approach which bare its fruits in the periodic case. The sole case that could have been treated so far by the form factor method corresponds to 
the edge spin-spin dynamical autocorrelation function \cite{GrijalvaDeNArdisTerrasXXZBoundaryMagnetisation}. 
Finally, some progress was achieved in describing the  XXZ chain subject to 
open boundary conditions at finite temperature: the surface free energy for the case of so-called diagonal boundary fields was characterised in the works
\cite{BortzFrahmGohmannSurfaceFreeEnergy,KozPozsgaySurfaceFreeEnergyBoundaryXXZ}
while the one associated with general boundary fields in the work \cite{PozsgayRakosBoundaryFreeEnergyGeneralBCXXZChain}.
However, so far, no closed expressions for the thermal correlation functions exist
in the case of open boundary conditions.

This paper aims at filling this gap. More precisely, we establish a setting allowing one to obtain thermal form factor expansions for multi-point correlation functions in the 
XXZ chain subject to diagonal boundary fields. As in the periodic case, these expansions fully factorise the distance dependence of the correlation function and bare, in fact, several
structural similarities with those arising in the case of the periodic chain. While, as already mentioned, the lack of translation invariance renders the zero-temperature
form factor expansions ineffective in the case of open boundary conditions, the key observation of this work is that the presence of finite temperature
allows one to bypass these limitations.  Indeed, as observed in \cite{BortzFrahmGohmannSurfaceFreeEnergy},
the quantum transfer matrix -- \textit{viz}. the auxiliary object naturally describing the thermodynamics of the \textit{periodic} XXZ chain --
appears as a key ingredient for computing the surface free energy of the  XXZ chain subject to diagonal boundary fields. The surface free energy 
is expressed as the Trotter limit of the average of a specific projector calculated in  respect to the dominant state of the quantum transfer matrix. 
The Trotter limit of that representation can be taken upon observing that the latter may be  recast \cite{KozPozsgaySurfaceFreeEnergyBoundaryXXZ} in terms of the partition functions of the 
six-vertex model with reflecting ends which admits a determinant representation \cite{TsuchiyaPartitionFunctWithReflecEnd}. 
In the present work, we push further this construction by conforming it to the computation of multi-point correlation functions. 
The connection between the periodic and open boundary conditions for finite temperature quantum integrable models unravelled in  \cite{BortzFrahmGohmannSurfaceFreeEnergy}
allows us to build on the concept of thermal form factor expansions \cite{KozDugaveGohmannThermaxFormFactorsXXZ,KozDugaveGohmannThermaxFormFactorsXXZOffTransverseFunctions} 
and the projector representation of  \cite{BortzFrahmGohmannSurfaceFreeEnergy} so as to set up thermal boundary form factor expansions in  XXZ chain subject to diagonal boundary fields. 
This constitutes the main result of this work.

The paper is organised as follows. 
In Section \ref{Section generalites}, the description of the open XXZ chain at finite temperature $T$ within the Trotter approximation method is briefly recalled, and this setting is applied to the derivation of a finite Trotter approximant 
for the thermal multi-point correlation functions in the open XXZ chain.
In Section \ref{Section correlateur a trotter fini}, we reorganise the result in the form of a Trotter limit of a finite 
Trotter number thermal form factor expansion. 
In Section \ref{Section limite de Trotter} is developed a scheme, following \cite{KozDugaveGohmannThermaxFormFactorsXXZ,KozPozsgaySurfaceFreeEnergyBoundaryXXZ},
for taking the Trotter limit in these expressions. We more particularly focus there on the case of the longitudinal spin one point function, the calculation of the Trotter limit for the multi-point correlation functions being a trivial generalisation thereof.
The result of this process for the one-point function at infinite Trotter limit, \textit{viz}. a closed expression for this one-point function at finite temperature, is finally presented in Section \ref{sec-result}.

\section{Multi-point correlation functions at finite temperature: setting of the problem}
\label{Section generalites}

We consider in this paper the XXZ spin-$\tf{1}{2}$ open chain with longitudinal boundary fields $h_-$ and $h_+$, and in a uniform external magnetic field $h$.
The corresponding Hamiltonian is
\beq
\label{Hh}
  \op{H}_h \; = \; \op{H} - \f{h}{ 2} \sul{k=1}{L}\sg_k^z \;, 
\enq
with
\beq
  \op{H}= J \sum_{m=1}^{L-1} \Big\{ \sigma^x_m \,\sigma^x_{m+1} +
  \sigma^y_m\,\sigma^y_{m+1} + \Delta \,(\sigma^z_m\,\sigma^z_{m+1}+ \e{id} )\Big\}
  + 
  h_-\, \sg^z_1 +  
  h_+\,\sg^z_L + c\, . 
\label{ecriture Hamiltonien XXZ ac bord}
\enq
This Hamiltonian acts on the quantum space of states $\mf{h}_{XXZ} = \otimes_{p=1}^{L}\mf{h}_p$, where the local quantum space  $\mf{h}_a$ associated with the $a^{\e{th}}$ site is isomorphic to $\Cx^2$.
Here $J$ is a global coupling constant, $\Delta$ is some anisotropy parameter along the $z$-direction, $\sigma^{x,y,z}$ are the Pauli matrices
and given an operator $\op{O}\in \e{End}(\Cx^2)$, $\op{O}_{m}\in \e{End}(\mf{h}_{XXZ})$ acts as $\op{O}$ on the tensor component $\mf{h}_m$
and as the identity on all the other tensor components.
In the following, we shall parameterise the anisotropy parameters and boundary fields in terms of three parameters $\zeta$ and $\xi_\pm$ as
\beq
\Delta=\cos (\zeta), \qquad h_\pm^z= J\, \sinh (-\i\zeta) \, \coth\xi_\pm,
\label{choice-parametr}
\enq
and fix for convenience the constant $c$ in \eqref{ecriture Hamiltonien XXZ ac bord} to be equal to 
\beq
c=J\,\f{\cos(2\zeta)}{\cos(\zeta)}  \; .
\enq
Although the parameterisation \eqref{choice-parametr} is most suited for the study of the massless regime $-1<\cos \zeta < 1$, we stress that the adaptation
of our handlings to the other regimes of the couplings is straightforward. 

Given $r$ local operators $\op{O}^{(1)}_{m_1+1}, \dots, \op{O}^{(r)}_{m_r+1}$, our aim is to compute the thermal average at temperature $T$,
\beq
\label{def-corr-L}
\mathbb{E}_{L;T}\Big[\op{O}^{(1)}_{m_1+1}  \cdots  \op{O}^{(r)}_{m_r+1}   \Big] \; = \; \e{tr}_{ \mf{h}_{XXZ} } \bigg\{ \op{O}^{(1)}_{m_1+1}  \dots  \op{O}^{(r)}_{m_r+1}  \ex{-\f{ \op{H}_h }{T} }  \bigg\}   \cdot Z_L^{-1},
\enq
in which $Z_L$ is the partition function of the model: 
\beq
\label{def-ZL}
Z_L \; = \; \e{tr}_{ \mf{h}_{XXZ} } \Big[ \ex{-\f{ \op{H}_h }{T} } \Big]\, . 
\enq
More precisely, we aim at computing the thermodynamic limit of \eqref{def-corr-L}:
\beq
\label{def-corr}
\big< \op{O}^{(1)}_{m_1+1}  \cdots  \op{O}^{(r)}_{m_r+1} \big>_{T} \; = \;  
\lim_{L\tend + \infty} \mathbb{E}_{L;T}\Big[\op{O}^{(1)}_{m_1+1}  \cdots  \op{O}^{(r)}_{m_r+1}   \Big]\, .
\enq

\subsection{A Trotter approximant of the partition function}

The  Hamiltonian \eqref{Hh} can be diagonalised by means of the boundary version of the algebraic Bethe Ansatz approach \cite{SklyaninABAopenmodels}. For this, one needs to introduce two monodromy matrices,
\beq
T_a(\la) \,  = \,  R_{aL}(\la-\xi_L)\dots R_{a1}(\la-\xi_1) \qquad \e{and} \qquad 
\wh{T}_a(\la) \,  = \, R_{1a}(\la+\xi_1) \dots R_{La}(\la+\xi_L) \;. 
\enq
There $\xi_k$ represent inhomogeneity parameters, the roman index $a$ refers 
to an auxiliary two-dimensional space whereas the indices $1,\dots,L$ refer
to the various quantum spaces $\mf{h}_1,\dots, \mf{h}_L$ arising in the tensor product decomposition of the model's Hilbert space $\mf{h}_{XXZ} = \otimes_{p=1}^{L}\mf{h}_p$.  
The above monodromy matrices are built from the six-vertex R-matrix taken in the polynomial normalisation:
\beq
\op{R}(\la)  =  \f{1}{ \sinh(-\i\zeta) } \left( \ba{cccc}  \sinh(\la-\i\zeta) &  0 &0 &0 \\
							0 & \sinh(\la)& \sinh(-\i\zeta) & 0 \\
								0 & \sinh(-\i\zeta)& \sinh(\la) & 0 \\
							0 & 0 & 0 & \sinh(\la-\i\zeta) \ea  \right) \;.
\enq
Further, one also needs to introduce the diagonal solutions of the reflection equations that have been first found by Cherednik \cite{CherednikReflectionEquationFactorisabilityOfScattering},
\beq
K^{\pm}_a(\la) = K_a\Big( \la -\i\tfrac{\zeta}{2} \mp \i\tfrac{\zeta}{2} ; \xi_{\pm}  \Big) \qquad \e{with} \quad 
K_a(\la;\xi) = \left(  \ba{cc} \sinh(\la+\xi) & 0 \\ 0 & \sinh(\xi-\la) \ea \right)_{[a]} \;.
\label{Kdef}
\enq
 These $K$-matrices can be checked to satisfy 
\beq
\e{tr}_a\big[ K^+_a(0) \big] = 2 \sinh(\xi_+) \cos(\zeta) \qquad \e{and}  \qquad 
\e{tr}_a\big[ K^-_a(0) \big] = 2 \sinh(\xi_-)  \;.
\label{ecriture traces matrices K}
\enq

Then, the model's transfer matrix takes the form 
\beq
\bs{\tau}(\la)  \; = \;  \e{tr}_{\mf{h}_a}\Big[  K^+_a(\la) T_a(\la) K^-_a(\la) \wh{T}_a(\la) \Big]   \;. 
\label{ecriture boundary transfer matrix}
\enq
One can show \cite{CherednikReflectionEquationFactorisabilityOfScattering} that, 
in the homogeneous limit ($\xi_k=0$, $k=1,\dots,L$), $\bs{\tau}(\la)$ enjoys  the properties 
\beq
{\bs{\tau}(0)}_{\mid \xi_a=0}  = \f{ \e{tr}_a\big[ K^+_a(0) \big] \e{tr}_a\big[ K^-_a(0) \big] }{2}   \op{id}
\qquad \e{and} \qquad 
\op{H} = \f{J \sinh(-\i \zeta) }{ {\bs{\tau}(0)}_{\mid \xi_a=0} } \f{\dd }{\dd \la} \bs{\tau} (\la) _{\mid_{\la=0,  \xi_a=0}} \;. 
\enq
There $\op{id}$ stands for the identity operator on $\mf{h}_{XXZ}$ and $\op{H} $ 
is given by \eqref{ecriture Hamiltonien XXZ ac bord}. 
As a consequence, 
\beq
\Big( \Big[ \bs{\tau}(-\tfrac{\be}{N} ) \cdot \bs{\tau}^{-1}(0) \Big]_{\mid \xi_a=0} \Big)^N= \ex{ - \f{ \op{H} }{T} } \cdot 
\Big(1 + \e{O}\big( N^{-1} \big) \Big) \qquad \e{with} \qquad \be = \f{ J\sinh(-\i\zeta) }{ T }.
\enq
This leads to a Trotter limit-based representation for the partition function \eqref{def-ZL}
of the open XXZ spin-$\tf{1}{2}$ chain in a uniform external magnetic field:
\beq
Z_L =  \lim_{N\tend +\infty}  \e{tr}_{\mf{h}_{XXZ} } \bigg\{ 
	\Big[ \bs{\tau}\big(-\tfrac{\be}{N} \big) \cdot \bs{\tau}^{-1}(0)  \Big]^N   \pl{a=1}{L} \ex{ \f{h}{2T}\sg_a^z } \bigg\}_{\mid \xi_a=0} \; . 
\label{ecriture fction partition comme trace}
\enq

\subsection{A Trotter approximant for multi-point functions}

The above approach also allows one to obtain a Trotter approximant of the thermal average \eqref{def-corr-L}:
\begin{align}
\mathbb{E}_{L;T}\Big[\op{O}^{(1)}_{m_1+1}  \cdots  \op{O}^{(r)}_{m_r+1}   \Big] \cdot Z_L \; &= \; \e{tr}_{ \mf{h}_{XXZ} } \bigg\{ \op{O}^{(1)}_{m_1+1}  \dots  \op{O}^{(r)}_{m_r+1}  \ex{-\f{ \op{H}_h }{T} }  \bigg\}  \nonumber\\
\; &= \;  \lim_{N\tend +\infty} \e{tr}_{ \mf{h}_{XXZ} } \bigg\{ \op{O}^{(1)}_{m_1+1}  \cdots  \op{O}^{(r)}_{m_r+1}   \cdot \bs{\tau}^{N}\big(-\tfrac{\be}{N} \big) \cdot \bs{\tau}^{-N}(0)   \cdot  \pl{a=1}{L} \ex{ \f{h}{2T}\sg_a^z }   \bigg\}_{\mid \xi_a=0}    \;.  
\end{align}
This may be recast in terms of quantities naturally associated with the quantum transfer matrix. 
For doing so, one needs to rely on  the representation, first observed in \cite{BortzFrahmGohmannSurfaceFreeEnergy},
\beq
\bs{\tau}(\la)  \; = \;  \e{tr}_{\mf{h}_a\otimes \mf{h}_b }\Big[  \op{P}_{ab}(\la)  T_b^{\mf{t}_b}(\la) \wh{T}_a(\la) \Big]  \qquad \e{with} \qquad 
\op{P}_{ab}(\la) = K_a^+(\la) \, \mc{P}_{\!\! ab}^{\mf{t}_a}  \,  K_a^{-}(\la)   \;,
\label{ecriture boundary transfer matrix representation BFG}
\enq
in which $\mc{P}_{\!\! ab}$ is the permutation operator on $\mf{h}_a\otimes \mf{h}_b$.
Then direct algebra leads to 
\bem
 \op{O}^{(1)}_{m_1+1}  \cdots  \op{O}^{(r)}_{m_r+1}   \cdot \bs{\tau}^{N}\big(-\tfrac{\be}{N} \big)  \pl{a=1}{L} \ex{ \f{h}{2T}\sg_a^z } \; = \; 
  \e{tr}_{ \mf{h}_{\mf{q}} }\Big[ \Pi_{\mf{q}}\big(-\tfrac{\be}{N} \big) \, \op{T}_{\mf{q};1}(\xi_1)\cdots \op{T}_{\mf{q};m_1}(\xi_{m_1}) \,  \op{O}^{(1)}_{m_1+1} \,  \op{T}_{\mf{q};m_1+1}(\xi_{m_1+1}) \cdots  \\
 \cdots   \op{O}^{(r)}_{m_r+1}\,  \op{T}_{\mf{q};m_r+1}(\xi_{m_r+1})\, \op{T}_{\mf{q};m_r+2}(\xi_{m_r+2}) \cdots \op{T}_{\mf{q};L}(\xi_{L}) \Big]
\label{equation pour rep fct multipts}
\end{multline}
in which $\mf{h}_{\mf{q}} =\otimes_{k=1}^{2N}\mf{h}_{a_k}$, $\mf{h}_{a_k}\simeq \Cx^2$, and $\op{T}_{\mf{q};k}$ is the quantum monodromy matrix with auxiliary space $\mf{h}_k$:
\beq
\label{def-QMM}
\op{T}_{\mf{q};k}(\xi) \,  = \,  R^{\mf{t}_{a_{2N}}}_{a_{2N} k }( - \xi -  \tfrac{\be}{N} )\, R_{k a_{2N-1}}(\xi - \tfrac{\be}{N} ) \cdots
R_{a_{2} k}^{ \mf{t}_{a_2} }(-\xi - \tfrac{\be}{N} )\, R_{k a_{1}}(\xi - \tfrac{\be}{N} )\, \ex{\f{h}{2T}  \sg^z_k } 
 \; = \; \pa{\ba{cc} \op{A}(\la) & \op{B}(\la) \\ \op{C}(\la) & \op{D}(\la) \ea  }_{ [k] } \;. 
\enq
Finally, $\Pi_{\mf{q}}(\la)$ is a rank one matrix defined as 
\beq
 \Pi_{\mf{q}}\big( \la \big)  \, = \, \op{P}_{a_1a_2}(\la)\dots \op{P}_{a_{2N-1}a_{2N}}(\la) \;,
\enq
and which can be factorised as 
\beq
\Pi_{\mf{q}}\big( \la \big) \; = \;   K_{a_1}^+(\la) \cdots  K_{a_{2N-1}}^+(\la) \, \bs{\op{v}} \cdot \bs{\op{v}}^{\mf{t}}   K_{a_{2N-1}}^-(\la)  \cdots K_{a_1}^-(\la) \, ,
\enq
where 
\beq
\bs{\op{v}} \, = \, \Big(\bs{e}_1\otimes \bs{e}_1  + \bs{e}_2\otimes \bs{e}_2   \Big) \otimes \cdots \otimes  \Big(\bs{e}_1\otimes \bs{e}_1  + \bs{e}_2\otimes \bs{e}_2   \Big) \;,
\enq
$\bs{e}_1$ and $\bs{e}_2$ being the elements of the canonical basis of $\mathbb{C}^2\simeq \mf{h}_{a_k}$.
Note that, in the case where there are no operators in \eqref{equation pour rep fct multipts}, one simply obtains the representation obtained in \cite{BortzFrahmGohmannSurfaceFreeEnergy}:
\beq
  \bs{\tau}^{N}\big(-\tfrac{\be}{N} \big) \, \pl{a=1}{L} \ex{ \f{h}{2T}\sg_a^z } \; = \; 
  \e{tr}_{ \mf{h}_{\mf{q}} }\Big[ \Pi_{\mf{q}}\big(-\tfrac{\be}{N} \big) \, \op{T}_{\mf{q};1}(\xi_1)\cdots \op{T}_{\mf{q};L}(\xi_{L}) \Big] \;. 
\enq

Thus, upon taking the traces over the local quantum spaces $\mf{h}_a$ building up the model's Hilbert space $\mf{h}_{XXZ}$, one arrives to the below representation for the finite volume $L$
thermal average of a multi-point function:
\bem
\label{corr-trotter-L}
\mathbb{E}_{L;T}\Big[\op{O}^{(1)}_{m_1+1}  \cdots  \op{O}^{(r)}_{m_r+1}   \Big]   \; = \; \lim_{N\tend + \infty} \Bigg\{  
\bigg\{  \e{tr}_{ \mf{h}_{\mf{q} } } \Big[   \Pi_{\mf{q}}\big(-\tfrac{\be}{N} \big)\,  \Big(\op{t}_{\mf{q}}(0) \Big)^L \Big]  \bigg\}^{-1}\\
\times \e{tr}_{ \mf{h}_{\mf{q} } } \Big[  \Pi_{\mf{q}}\big(-\tfrac{\be}{N} \big) \cdot \Big( \op{t}_{\mf{q}}(0) \Big)^{m_1} \cdot \Xi^{(1)}  \cdot \Big( \op{t}_{\mf{q}}(0) \Big)^{m_2-m_1-1} \cdot \Xi^{(2)}  \cdots 
\Xi^{(r-1)}  \cdot \Big( \op{t}_{\mf{q}}(0) \Big)^{m_r-m_{r-1}-1} \cdot \Xi^{(r)} \cdot \Big( \op{t}_{\mf{q}}(0) \Big)^{L-m_r-1} \Big]
  \Bigg\} \;.  
\end{multline} 
Here $\op{t}_{\mf{q}}(0)$ stands for the quantum transfer matrix
\beq
\label{def-TQM}
\op{t}_{\mf{q}}(0) \, = \, \e{tr}_{\mf{h}_0}\Big[ \op{T}_{\mf{q};0}(0)   \Big]  \;,
\enq
and we have also defined
\beq
\Xi^{(k)}\, = \, \e{tr}_{\mf{h}_0}\Big[ \op{T}_{\mf{q};0}(0) \, \op{O}^{(k)}_0 \Big]   \;. 
\enq

To proceed further, \textit{viz}. take the thermodynamic limit of the thermal expectation \eqref{corr-trotter-L}, one needs to assume that  one may exchange the Trotter and the thermodynamic limits. 
This has been rigorously established, for $T$ sufficiently large, in the case of the quantum transfer matrix approach to the  periodic  chain in \cite{KozGohmannGoomaneeSuzukiRigorousApproachQTMForFreeEnergy}. 
Further, that work also established that, at least for $T$ large enough, $\op{t}_{\mf{q}}(0)$ admits a maximal in modulus, real Eigenvalue $\wh{\La}_{0;0}$. The associate dominant Eigenvector
is denoted by $\bs{\Psi}_{0}^{(0)}$. Thus, under the assumption of commutativity of the Trotter and thermodynamic limits, the thermal average of an $r$ point function projects onto 
the Trotter limit of an average of a string of operators: 
\bem
\label{corr-trotter}
\big< \op{O}^{(1)}_{m_1+1}  \cdots  \op{O}^{(r)}_{m_r+1} \big>_{T} \; \\
= \;  \lim_{N\tend + \infty} \left\{  
\f{  \Big(\bs{\Psi}_0^{(0)}, \Pi_{\mf{q}}\big(-\tfrac{\be}{N} \big) \,  \Big( \op{t}_{\mf{q}}(0) \Big)^{m_1} \, \Xi^{(1)}  \,
\Big( \op{t}_{\mf{q}}(0) \Big)^{m_2-m_1-1} \, \Xi^{(2)}  \cdots \Xi^{(r-1)} \, \Big( \op{t}_{\mf{q}}(0) \Big)^{m_r-m_{r-1}-1} \, \Xi^{(r)}\, \bs{\Psi}_{0}^{(0)}\Big) }
{  \Big(\bs{\Psi}_0^{(0)}, \Pi_{\mf{q}}\big(-\tfrac{\be}{N} \big) \, \bs{\Psi}_{0}^{(0)}\Big) \cdot \wh{\La}_{0;0}^{m_r+1}  }  \right\} \;. 
\end{multline}

The above representation constitutes the starting point of our derivation of the boundary thermal form factor expansion of the multi-point functions.

\section{The finite Trotter thermal form factor series expansion}
\label{Section correlateur a trotter fini}

In this section, we explicitly express the general multi-point correlation function \eqref{corr-trotter} as the Trotter limit of a sum over a product of finite Trotter thermal form factors --- the bulk quantum transfer matrix elements --- multiplied by some boundary factor which contains the whole information on the boundary fields. 
So as to write this form factor expansion, we need to make a short detour relatively to the spectrum of the quantum transfer matrix \eqref{def-TQM}.

\subsection{The algebraic Bethe Ansatz approach to the quantum transfer matrix}

Recall that  Eigenstates of the quantum transfer matrix \eqref{def-TQM}
can be constructed from the knowledge of the solutions to the following Bethe Ansatz equations, 
\beq
-1 =  (-1)^{s} \, \ex{-\f{h}{T} } \pl{k=1}{M} \bigg\{  \f{ \sinh(\i \zeta + \mu_k-\mu_p )  }{  \sinh(\i\zeta + \mu_p - \mu_k  )  }  \bigg\} \cdot 
 \bigg[  \f{ \sinh(\mu_p+\tf{\be}{N}+ \i \zeta)  \sinh( \mu_p- \tf{\be}{N})  }{  \sinh(\i\zeta - \mu_p + \tf{\be}{N} ) \sinh(\mu_p + \tf{\be}{N})  } \bigg]^N \; , 
\label{ecriture BAE QTM}
\enq
under the hypothesis that the roots are pairwise distinct, see \cite{KozGohmannGoomaneeSuzukiRigorousApproachQTMForFreeEnergy}. Here $s=N-M$ is called the spin. 
The Eigenstates associated with a solution $\{\mu_a\}_1^M$  will be written as $\bs{\Psi}\Big( \{\mu_a\}_1^M \Big)$. 
These satisfy
\beq
\op{t}_{\mf{q}}(0) \cdot \bs{\Psi}\Big( \{\mu_a\}_1^M \Big) \; = \;   \tau\big( 0 \mid \{\mu_a\}_1^{M} \big) \cdot  \bs{\Psi}\Big( \{\mu_a\}_1^M \Big)
\enq
in which 
\bem
 \tau\big( \xi \mid \{\mu_k\}_1^M \big) \, = \,  (-1)^N \ex{\f{h}{2T} }\pl{k=1}{M} \Bigg\{ \f{ \sinh(\xi-\mu_k + \i \zeta) }{ \sinh(\xi-\mu_k )  }  \Bigg\} \cdot 
\Bigg( \f{ \sinh(\xi + \tf{\be}{N} ) \sinh(\xi - \tf{\be}{N} -\i \zeta) }{ \sinh^2(-\i\zeta) } \Bigg)^N   \\ 
\, + \,  (-1)^N  \ex{-\f{h}{2T} } \pl{k=1}{M}  \Bigg\{ \f{ \sinh(\xi-\mu_k - \i \zeta) }{ \sinh(\xi-\mu_k ) }  \Bigg\} \cdot 
 \Bigg(  \f{ \sinh(\xi + \tf{\be}{N} +\i\zeta) \sinh(\xi - \tf{\be}{N})  }{ \sinh^2(-\i\zeta) }  \Bigg)^N  .
\label{ecriture valeur propre qtm}
\end{multline}
For convenience, we shall sometimes insist on the dependence on the magnetic field $h$ 
\begin{itemize}
 
 \item[i)] of the matrix entries of the quantum monodromy matrix $\op{A}_{h}, \op{D}_{h}, ....$, 
 
 \item[ii)]of the quantum transfer matrix $\op{t}_{\mf{q};h}$, 
 
  \item[iii)] of the Bethe roots $ \{\mu_a(h)\}_1^{M}$, 
 
 \item[iii)] of the Eigenvalues $\tau_{h}\big( 0 \mid \{\mu_a\}_1^{M} \big)$.

\end{itemize}

It has been rigorously shown in \cite{KozGohmannGoomaneeSuzukiRigorousApproachQTMForFreeEnergy} that, for $T$ large enough, uniformly in $N$, $\op{t}_{\mf{q}}(0)$
has a dominant Eigenvalue $\wh{\La}_{0;0}$ which can be built from a solution $\{\la_a\}_1^N$ to \eqref{ecriture BAE QTM} with $M=N$. As already discussed, the associated
Eigenvector $\bs{\Psi}\Big( \{\la_a\}_1^M \Big)$ is denoted $\bs{\Psi}^{(0)}_0$ and called the dominant state of the quantum transfer matrix.
>From now on, $\{\la_a\}_1^N$ will always refer to the set of Bethe roots describing the dominant state.

In a given spin sector $s$ above the dominant state, \textit{viz}. corresponding to Eigenvectors parameterised by 
\beq
M=N-s
\enq
Bethe roots, one may denote, for short,  the Eigenvectors of the quantum transfer matrix as $\bs{\Psi}_k^{(s)}$. In this notation, 
$\bs{\Psi}_k^{(s)}$ is assumed to be a Bethe vector parameterised by the Bethe roots $\{\mu_a^{(k;s)}\}_1^{N-s}$. 
Then, one also introduces the notation $\wh{\La}_{k; s}$ for the associated Eigenvalues, \textit{viz}. 
\beq
\label{eigenvectors-notation}
\op{t}_{\mf{q}}(0) \cdot \bs{\Psi}_k^{(s)}  \, = \, \wh{\La}_{k; s} \cdot \bs{\Psi}_k^{(s)} \;. 
\enq
Clearly, in order for the above to be well defined, one needs that the quantum transfer matrix is diagonalisable and that 
the Bethe Ansatz construction provides one with its complete set of Eigenvectors. 
We shall not dwell on the diagonalisability/completeness issues here and take them for granted or simply irrelevant for
the limits considered.

\subsection{The general case of multi-point functions}

Upon inserting a sum over a complete set of Eigenstates $\bs{\Psi}_{k_a}^{(s_a)}$
in front of each of the operators $\Xi^{(k)}$ in the expression \eqref{corr-trotter}, one obtains a form factor representation for the multi-point correlation functions:
\begin{multline}
\big< \op{O}^{(1)}_{m_1+1}  \cdots  \op{O}^{(r)}_{m_r+1} \big>_{T} \; 
\\
= \;  \lim_{N\tend + \infty} \left\{  \sul{  \substack{ k_a \in \mf{D}_a  \\ a = 1, \dots, r} }{}
%
\wh{\La}_{k_1; s_1}^{m_1} \pl{a=2}{r} \Big\{ \wh{\La}_{k_a;s_a}^{m_a-m_{a-1}-1} \Big\} \, \wh{\La}_{0;0}^{-m_r-1}\cdot
\f{\Big(\bs{\Psi}_0^{(0)}, \Pi_{\mf{q}}\big(-\tfrac{\be}{N} \big)  \bs{\Psi}_{k_1}^{(s_1)}\Big)}{\Big(\bs{\Psi}_0^{(0)}, \Pi_{\mf{q}}\big(-\tfrac{\be}{N} \big)  \bs{\Psi}_{0}^{(0)}\Big)}
\cdot \pl{a=1}{r} \f{   \Big(\bs{\Psi}_{k_a}^{(s_a)},\Xi^{(a)}     \bs{\Psi}_{k_{a+1}}^{(s_{a+1})}\Big)   }{   \Big(\bs{\Psi}_{k_a}^{(s_a)},  \bs{\Psi}_{k_a}^{(s_a)}\Big)   }  \right\} \;. 
\label{ecriture limite Trotter fct r pts tempe finie bord-1}
\end{multline}
Here we have used the notations \eqref{eigenvectors-notation}, and we agree upon $k_{r+1}=s_{r+1}=0$. Also,  the spins $s_a$ of the inserted Eigenstates are to be taken such that 
\beq
\label{matrix-el-QTM}
\Big(\bs{\Psi}_{k_a}^{(s_a)},\Xi^{(a)}     \bs{\Psi}_{k_{a+1}}^{(s_{a+1})}\Big) \; \not= \; 0. 
\enq
In \eqref{ecriture limite Trotter fct r pts tempe finie bord}, the summations run over appropriate subsets $\mf{D}_a$ for the $k_a$'s
which are compatible with the spin constraints. Also, note that the constraint \eqref{matrix-el-QTM} means that we focus on a class of local operators
$\op{O}^{(a)}\in \e{End}(\Cx^2)$ having a definite spin (\textit{i.e.} $\e{id}, \sg^z, \sg^{\pm}$). More general operators are of course
allowed but then one needs to sum in \eqref{ecriture limite Trotter fct r pts tempe finie bord-1} also over the spins $s_a$ on the
intermediate states.

We would like to underline that the matrix elements \eqref{matrix-el-QTM}
are the bulk quantum transfer matrix elements, for which there exist a convenient determinant representation at finite Trotter number $N$, see \cite{SlavnovScalarProductsXXZ}. Their Trotter limit can be studied similarly as in \cite{KozDugaveGohmannThermaxFormFactorsXXZ}.
Quite remarkably, the whole information on the boundary field is contained, for each term of the sum, in a single factor, the ratio of matrix elements
\begin{equation}\label{boundary-factor}
   \f{\Big(\bs{\Psi}_0^{(0)}, \Pi_{\mf{q}}\big(-\tfrac{\be}{N} \big)  \bs{\Psi}_{k_1}^{(s_1)}\Big)}{\Big(\bs{\Psi}_0^{(0)}, \Pi_{\mf{q}}\big(-\tfrac{\be}{N} \big)  \bs{\Psi}_{0}^{(0)}\Big)}.
\end{equation}
The latter can also be represented, at finite Trotter number $N$ in terms of a determinant of size $N$, as recalled here below.

Given two Bethe vectors $\bs{\Psi}_0^{(0)}$ and $\bs{\Psi}_k^{(0)}$ parameterised by the roots $\{\la_a\}_1^N$, resp. $\{\mu_a^{(k;0)}\}_1^N$, it was indeed established in \cite{KozPozsgaySurfaceFreeEnergyBoundaryXXZ}
that one has 
\beq
\Big(\bs{\Psi}_0^{(0)}, \Pi_{\mf{q}}\big(-\tfrac{\be}{N} \big)  \bs{\Psi}_{k}^{(0)}\Big) \; = \; \mc{F}^{(+)}\big( \{\la_a\}_1^N \big) \cdot  \mc{F}^{(-)}\big( \{\mu_a^{(k;0)} \}_1^N \big)
\enq
in which 
\beq
\label{def-F-}
 \mc{F}^{(-)}\big( \{\mu_a\}_1^N \big)\; = \; \ex{ - \f{N h}{2T} } \cdot \mc{Z}_N\Big( \{ - \tfrac{\be}{N}  \}_1^N ; \{\mu_a\}_1^N; \xi_- \Big)
\enq
is expressed in terms of the partition function of the six-vertex model with reflecting ends \cite{TsuchiyaPartitionFunctWithReflecEnd}:
\bem
\mc{Z}_N\Big( \{\xi_a \}_1^N ; \{\mu_a\}_1^N; \xi_- \Big) \,  =  \,   
\f{  \pl{a,b=1}{N} \pl{\eps=\pm}{} \Big\{ \sinh(\xi_a+\eps \mu_b)  \sinh(\xi_a-\i\zeta + \eps \mu_b) \Big\}  }
{   \pl{a<b}{N}  \Big\{ \sinh(\xi_a-\xi_b) \sinh(\xi_a+\xi_b-\i\zeta)  \pl{\eps=\pm}{} \sinh(\mu_b+\eps\mu_a)  \Big\} } \\ 
\times   \det_{N}\Bigg[ \f{  \sinh(-\i\zeta)   \sinh(\xi_- + \mu_b)  \sinh(2\xi_a)  }{  \pl{\eps=\pm}{} \sinh(\xi_a-\i\zeta+ \eps \mu_b)\sinh(\xi_a+\eps\mu_b)  } \Bigg] \;.
\label{ecriture Tsuchiya determinant}
\end{multline}
A similar expression can be written for  $\mc{F}^{(+)}\big( \{\la_a\}_1^N \big)$ (see \cite{KozPozsgaySurfaceFreeEnergyBoundaryXXZ}), but it is clear that this factor disappears when considering the ratio \eqref{boundary-factor} and therefore will not play any role in the following, so that we omit to recall it here. 
Also, due to symmetry reasons, it holds that 
\beq
\Big(\bs{\Psi}_0^{(0)}, \Pi_{\mf{q}}\big(-\tfrac{\be}{N} \big)  \bs{\Psi}_{k}^{(s)}\Big) \; = \; 0
\label{ecriture condition nullite spin non nul du projecteur bord}
\enq
whenever $s\not=0$.

Hence the  form factor representation \eqref{ecriture limite Trotter fct r pts tempe finie bord-1} can be rewritten as
\begin{multline}
\big< \op{O}^{(1)}_{m_1+1}  \cdots  \op{O}^{(r)}_{m_r+1} \big>_{T} \; \\
= \;  \lim_{N\tend + \infty} \left\{  \sul{  \substack{ k_a \in \mf{D}_a  \\ a = 1, \dots, r} }{}
  \wh{\La}_{k_1; 0}^{m_1} \pl{a=2}{r} \Big\{ \wh{\La}_{k_a;s_a}^{m_a-m_{a-1}-1} \Big\} \,  \wh{\La}_{0;0}^{-m_r-1} \cdot \f{  \mc{F}^{(-)}\big( \{\mu_a^{(k_1;0)}\}_1^N \big) }{   \mc{F}^{(-)}\big( \{\la_a\}_1^N \big) }
\cdot \pl{a=1}{r} \f{   \Big(\bs{\Psi}_{k_a}^{(s_a)},\Xi^{(a)}     \bs{\Psi}_{k_{a+1}}^{(s_{a+1})}\Big)   }{   \Big(\bs{\Psi}_{k_a}^{(s_a)},  \bs{\Psi}_{k_a}^{(s_a)}\Big)   }  \right\} \;,
\label{ecriture limite Trotter fct r pts tempe finie bord}
\end{multline}
with here $s_1=0$ due to \eqref{ecriture condition nullite spin non nul du projecteur bord}. 
We remind that $\{\la_a\}_1^N $ stands for the set of Bethe roots describing the dominant Eigenstate of the quantum transfer matrix. 
Also, we used explicitly that the number of Bethe roots describing the first inserted excited state $\bs{\Psi}_{k_1}^{(0)}$ is exactly 
$N$.

\subsection{The case of one-point functions}

We now specialise the above framework to the interesting particular case of the one-point functions,  \textit{viz}.  the thermal average of some local operator at distance $m$ from the boundary. By symmetry, the sole non-trivial one-point functions that are non-zero are the thermal expectation of the operator $\sg_{m+1}^{z}$.
For the sake of simplicity, from now on, we denote by  $\{ \mu_a (h^{\prime}) \}_1^{N}$ a set of solutions to the Bethe Ansatz equations \eqref{ecriture BAE QTM} at external magnetic field $h^{\prime}$. 
With this being settled, one may specialise the previous results as follows.
\begin{lemme}
It holds
\beq
\label{expr-1pt-fct}
\big< \sg_{m+1}^{z} \big>_{T} \; = \;  \lim_{N\tend + \infty}  \Big\{  2 T  \Dp{h^{\prime}} \mf{D}_m \mc{Q}_N(h^{\prime},m) \Big\}_{\mid h^{\prime}=h}
\enq
 where $ \mf{D}_m u=u_{m+1}-u_m$ and
\beq
\label{expr-QN}
\mc{Q}_N(h^{\prime},m) \; = \;   \sul{  \{ \mu_a (h^{\prime}) \}_1^N  }{}
\ex{  \f{N (h'-h)}{2T} } \,\f{  \mc{F}^{(-)}\big( \{ \mu_a(h^{\prime}) \}_1^N \big) }{   \mc{F}^{(-)}\big( \{\la_a(h) \}_1^N \big) }  
\cdot \f{   \Big( \bs{\Psi}( \{ \mu_a(h^{\prime}) \}_1^N ) ,   \bs{\Psi}( \{ \la_a(h) \}_1^N )  \Big)   }
{       \Big(\bs{\Psi}\big( \{ \mu_a(h^{\prime}) \}_1^N \big),  \bs{\Psi}\big( \{ \mu_a(h^{\prime}) \}_1^N \big) \Big)   } 
\cdot \Bigg( \f{ \tau_{h^{\prime}}\big( 0 \mid \{\mu_a(h^{\prime}) \}_1^{N} \big)  }{ \tau_{h}\big( 0 \mid \{\la_a(h) \}_1^{N} \big)  } \Bigg)^m \;. 
\enq
 Above, the sum runs over all solutions to the Bethe Ansatz equations \eqref{ecriture BAE QTM}

\end{lemme}

\Proof

Making the expansion \eqref{ecriture limite Trotter fct r pts tempe finie bord} explicit, one gets 
\beq
\big< \sg_{m+1}^{z} \big>_{T}  =    \lim_{N\tend + \infty} \left\{  \sul{  \{ \mu_a(h) \}_1^N  }{}
\f{  \mc{F}^{(-)}\big( \{ \mu_a(h) \}_1^N \big) }{   \mc{F}^{(-)}\big( \{\la_a(h)\}_1^N \big) }  
\cdot \f{   \Big( \bs{\Psi}( \{ \mu_a(h) \}_1^N ) , (\op{A}-\op{D})(0)  \cdot \bs{\Psi}( \{ \la_a(h) \}_1^N )  \Big)   }
{       \tau_h\big( 0 \mid \{\la_a(h)\}_1^{N} \big) \cdot  \Big(\bs{\Psi}\big( \{ \mu_a(h) \}_1^N \big),  \bs{\Psi}\big( \{ \mu_a(h) \}_1^N \big) \Big)   } 
\cdot \Bigg( \f{ \tau_h\big( 0 \mid \{\mu_a(h)\}_1^{N} \big)  }{ \tau_h\big( 0 \mid \{\la_a(h)\}_1^{N} \big)  } \Bigg)^m  \right\} . 
\nonumber 
\enq
Observe that it holds
\beq
 \Dp{h^{\prime}} \op{t}_{\mf{q}; h^{\prime}}(0)_{\mid h^{\prime}=h} \; = \; \f{1}{2T} \Big( \op{A}_h(0) - \op{D}_h(0)  \Big)  \;. 
\enq
 From there, it readily follows that 
\beq
\f{ \Big( \bs{\Psi}( \{\la_a(h)\}_1^N ) ,   (\op{A}_h-\op{D}_h)(0)  \bs{\Psi}( \{\mu_a(h)\}_1^N )\Big)   }
		{    \tau_h\big( 0 \mid \{ \la_a(h) \}_1^N \big)  }   = 
2T \f{ \Dp{} }{ \Dp{} h^{\prime} }  \Bigg\{  \;   \bigg( \f{ \tau_{h^{\prime}}\big( 0 \mid \{\mu_a(h^{\prime})\}_1^N \big) }{ \tau_h\big( 0 \mid \{\la_a(h)\}_1^N \big) }  -1 \bigg)
 \cdot  \Big( \bs{\Psi}( \{\mu_a(h^{\prime})\}_1^N ) , \bs{\Psi}( \{\la_a(h)\}_1^N ) \Big)  \Bigg\}_{\mid h=h^{\prime}} \;. 
\nonumber
\enq
Then, by using the discrete lattice derivative $ \mf{D}_m u=u_{m+1}-u_m$, one gets that 
\bem
\f{  \mc{F}^{(-)}\big( \{ \mu_a(h) \}_1^N \big) }{   \mc{F}^{(-)}\big( \{\la_a(h)\}_1^N \big) }  
\cdot \f{   \Big( \bs{\Psi}( \{ \mu_a(h) \}_1^N ) , (\op{A}_h-\op{D}_h)(0)  \cdot \bs{\Psi}( \{ \la_a(h) \}_1^N )  \Big)   }
{       \tau_h\big( 0 \mid \{\la_a(h)\}_1^{N} \big) \cdot  \Big(\bs{\Psi}\big( \{ \mu_a(h) \}_1^N \big),  \bs{\Psi}\big( \{ \mu_a(h) \}_1^N \big) \Big)   } 
\cdot \Bigg( \f{ \tau_h\big( 0 \mid \{ \mu_a(h) \}_1^{N} \big)  }{ \tau_h\big( 0 \mid \{ \la_a(h) \}_1^{N} \big)  } \Bigg)^m   \\
= 2 T  \Dp{h^{\prime}} \mf{D}_m\Bigg\{ \ex{  \f{N (h'-h)}{2T} } \,\f{  \mc{F}^{(-)}\big( \{ \mu_a(h^{\prime}) \}_1^N \big) }{   \mc{F}^{(-)}\big( \{\la_a(h) \}_1^N \big) }  
\cdot \f{   \Big( \bs{\Psi}( \{ \mu_a(h^{\prime}) \}_1^N ) ,   \bs{\Psi}( \{ \la_a(h) \}_1^N )  \Big)   }
{       \Big(\bs{\Psi}\big( \{ \mu_a(h^{\prime}) \}_1^N \big),  \bs{\Psi}\big( \{ \mu_a(h^{\prime}) \}_1^N \big) \Big)   } 
\cdot \Bigg( \f{ \tau_{h'}\big( 0 \mid \{\mu_a(h^{\prime}) \}_1^{N} \big)  }{ \tau_{h}\big( 0 \mid \{\la_a(h) \}_1^{N} \big)  } \Bigg)^m  \Bigg\} \;. 
\end{multline}
The claim follows upon putting the formulae together. \qed

 \section{Taking the Trotter limit}
\label{Section limite de Trotter}
 
 In order to obtain a closed expression for the multi-point correlation functions at finite $T$ one should still show that the Trotter limit can be taken on the level of \eqref{ecriture limite Trotter fct r pts tempe finie bord}. 
This demands to extract the large-$N$ behaviour of the matrix elements ratios 
\beq
\f{   \Big(\bs{\Psi}_{k_a}^{(s_a)},\Xi^{(a)}     \bs{\Psi}_{k_{a+1}}^{(s_{a+1})}\Big)   }{   \Big(\bs{\Psi}_{k_a}^{(s_a)},  \bs{\Psi}_{k_a}^{(s_a)}\Big)   }\, ,
\label{ratio elements matrices}
\enq
as well as the one of the boundary factor given by the ratios of partition functions of the six-vertex model with reflecting ends 
\beq
\label{boundary factor}
\mc{F}_{\mc{B}}\big( \{ \mu_a(h^{\prime}) \}_1^N ; \{\lambda_a(h)\}_1^N; \xi_- \big) 
=\ex{  \f{N (h'-h)}{2T} } \,\f{  \mc{F}^{(-)}\big( \{ \mu_a(h^{\prime}) \}_1^N \big) }{   \mc{F}^{(-)}\big( \{\la_a(h) \}_1^N \big) } 
\; = \;  \frac{\mc{Z}_N\Big( \{ - \tfrac{\be}{N}  \}_1^N ; \{\mu_a(h')\}_1^N; \xi_- \Big) }{\mc{Z}_N\Big( \{ - \tfrac{\be}{N}  \}_1^N ; \{\lambda_a(h)\}_1^N; \xi_- \Big) } \;. 
\enq
As already mentioned, the ratios \eqref{ratio elements matrices} can be expressed in terms of ratios of determinants whose size grows linearly with $N$ 
by use of the Gaudin and Slavnov representations \cite{KorepinNormBetheStates6-Vertex,SlavnovScalarProductsXXZ}. 
When $k_{a+1}=s_{a+1}=0$, their Trotter limit  was considered in \cite{KozDugaveGohmannThermaxFormFactorsXXZ} and, although cumbersome,  the generalisation of these considerations
to generic $k_{a+1}$ and $s_{a+1}$ is straightforward. The large-$N$ behaviour of  $\mc{F}^{(-)}\big( \{\la_a(h) \}_1^N \big)$, \textit{viz}. for the distribution of Bethe roots describing the dominant state of the quantum transfer matrix,
was also considered in  \cite{KozPozsgaySurfaceFreeEnergyBoundaryXXZ}.  Note that, for the consideration of the one-point function \eqref{expr-1pt-fct} or of the more general correlation function \eqref{ecriture limite Trotter fct r pts tempe finie bord}, we need here to consider the more general case involving the set of Bethe roots for an arbitrary excited state of the quantum transfer matrix.

The purpose of this section is to explain how one can reformulate the matrix elements \eqref{ratio elements matrices} and the boundary factor \eqref{boundary factor} in a convenient way for the consideration of the Trotter limit. As usual in the QTM approach, we use for that the reformulation of the QTM spectrum in terms of some non-linear integral equation for the associated counting function. We shall see that both quantities \eqref{ratio elements matrices} and \eqref{boundary factor} can be expressed as appropriate contour integrals involving this counting function.

In the following, we shall not consider the most general case and focus on the ratios arising in the study of one-point functions \eqref{expr-1pt-fct}-\eqref{expr-QN} since the latter already highlights 
all the technicalities arising in the computation of the Trotter limit.  
Also, for definiteness, we shall focus on the regime $\zeta\in ]0;\pi[$ ($-1<\Delta<1$).

 \subsection{The non-linear integral description of the spectrum of the quantum transfer matrix}

 To study the solutions to the Bethe equations and provide an exact characterisation of certain spectral properties of the quantum transfer matrix - see \cite{KozGohmannGoomaneeSuzukiRigorousApproachQTMForFreeEnergy} 
for a precise statement-, following \cite{DestriDeVegaAsymptoticAnalysisCountingFunctionAndFiniteSizeCorrectionsinTBAFirstpaper,KlumperNLIEfromQTMDescrThermoXYZOneUnknownFcton}, 
it is useful to introduce the counting function associated with a solution $\{\mu_a\}_1^M$ to the Bethe Ansatz equations 
\beq
\wh{\mf{a}}\big( \xi \mid \{\mu_a\}_1^M \big) \, = \, \ex{-\tfrac{h}{T}} (-1)^{s} \pl{k=1}{M} \bigg\{ \f{\sinh( \i\zeta - \xi + \mu_k) }{  \sinh( \i \zeta  + \xi - \mu_k) }  \bigg\}
\cdot  \bigg\{ \f{ \sinh(  \xi - \tf{ \be }{N} ) \sinh( \i\zeta +  \xi +\tf{ \be }{N} ) }{  \sinh(  \xi + \tf{ \be }{N} ) \sinh( \i\zeta - \xi + \tf{ \be }{N} )  }    \bigg\}^{N} \;. 
\label{definition fct auxiliaire a}
\enq

\subsubsection{The non-linear equation at finite Trotter number}

The main point is that the counting function can be expressed in an alternative way which allows one to study the Trotter limit and, subsequently, the low-$T$ limit efficiently. 
For that purpose, one first focuses on the solution describing the dominant state of the quantum transfer matrix. We remind that these roots are denoted by $\{\la_a\}_1^N$. 
Those roots were thoroughly characterised, on rigorous grounds, in \cite{KozGohmannGoomaneeSuzukiRigorousApproachQTMForFreeEnergy} for temperatures large enough.
Basically, one chooses a base contour $\msc{C}$ that encircles all the roots $\{\la_a\}_1^N$ as well as a neighbourhood of the origin, but not any other roots of $1+\wh{\mf{a}}\big( \xi \mid \{\la_a\}_1^N \big)$. 
By construction, the contour $\msc{C}$ is such that $1+\wh{\mf{a}}\big( \xi \mid \{\la_a\}_1^N \big)$ enjoys the zero monodromy condition relatively to it:
\beq
\label{monodromy-a}
0 = \Oint{  \msc{C} }{} \f{\dd \xi }{2\i\pi }  \f{  \wh{\mf{a}}^{\, \prime}\big( \xi \mid \{\la_a\}_1^N \big)  }{ 1+\wh{\mf{a}}\big( \xi \mid \{\la_a\}_1^N \big) } \;. 
\enq
We shall denote by $\msc{D}$ the compact domain such that $\Dp{}\msc{D}=\msc{C}$.

One may also construct an associated contour $\msc{C}_{s}$ relatively to the dominant Eigenstate of the quantum transfer matrix in the 
spin $s$ sector. From now on, for practical reasons, we shall restrict our analysis to the case $s=0$ which is directly relevant to the one-point function setting.

Then a sub-dominant Eigenstate of the quantum transfer matrix in the $s=0$ sector is characterised by the data:
\begin{itemize}
 
 \item a set $\wh{\mf{X}}$  gathering the positions of hole roots contained inside of $\msc{C}$;

 \item a set $\wh{\mc{Y}}$ gathering the positions of the particle roots contained outside of $\msc{C}$ within any $\i\pi$-periodic strip;

 \item the set of singular roots $\wh{\mc{Y}}_{\e{sg}}$ built out of  a subset of $\wh{\mc{Y}}$: $\wh{\mc{Y}}_{\e{sg}}=\big\{  y-\i \zeta_{\e{m}} \mf{s}_{2} \, : \, y \in \wh{\mc{Y}} \; \e{and} \;  y-\i \zeta_{\e{m}} \mf{s}_{2} \in \e{Int}(\msc{C}) \big\}$, where $\zeta_{\e{m}}=\e{min}(\zeta, \pi-\zeta)$ and $\mf{s}_{2}=\mathrm{sign}(\pi-2\zeta)$. 
 
\end{itemize}
While this is not essential for our calculations, we shall henceforth make the simplifying assumption that the roots 
building $\wh{\mc{Y}}$, $\wh{\mc{Y}}_{\e{sg}}$ and $\wh{\mf{X}}$ are all pairwise distinct and that we deal solely with excited states having no singular roots, 
\textit{viz} $\wh{\mc{Y}}_{\e{sg}}= \emptyset$. Treating the general case will not pose any problems but will definitely lead to cumbersome calculations
\symbolfootnote[2]{In particular, in case singular roots are present, one is in a situation where the functions $1+\mf{a}_{\mathbb{Y}}$ has additional poles inside of the contour
located at the singular roots. This gives rise to new contributions upon applying the product/sum $\hookrightarrow$ integral \textit{vs}.
$\tf{ \mf{a}_{\mathbb{Y}}^{\prime}  }{ \big( 1 + \mf{a}_{\mathbb{Y}} \big) }$ technique described later in the paper. Hence, this would change the fine details of the expression for the
infinite Trotter number limit of the bulk or boundary form factors appearing in \eqref{expr-QN}. We however do not expect that this would impact on the
overall properties of the site-$m$ magnetisation and, in particular, on its low-$T$ limit where one expects the appearance of a universal behaviour in the
large-distance regime. Moreover, as demonstrated rigorously in \cite{KozFaulmanGohmannLowTNLIERigourousAnalysisForQTMMasslessRegime}, singular
roots do not exist in the low-$T$ limit for $0\leq \De <1$ and there a very good indications that the statement also holds true for $-1<\De <0$.}.

Following the notations of \cite{KozGohmannGoomaneeSuzukiRigorousApproachQTMForFreeEnergy},
it is convenient to gather all the roots parameterising a sub-dominant state in terms of the formal difference of sets
\beq
\label{set-def}
\wh{ \mathbb{Y} } \, = \, \wh{\mc{Y}}  \ominus \wh{\mf{X}} \;. 
\enq
The main statement of the non-linear integral equation based approach to the spectrum of the quantum transfer matrix approach is that one may equivalently 
parameterise the counting function associated with an excited state $\{\mu_a\}_1^M$ in terms of the set $\wh{\mathbb{Y}}$ and of $s$. 
Henceforth, we shall thus denote the associated counting function as $\wh{\mf{a}}_{\mathbb{Y}}$ for short. 
In particular, he counting function associated with the dominant state is denoted $\wh{\mf{a}}_{\emptyset}$. Within this convention, 
the zero monodromy condition on the contour $\msc{C}$ translates itself into a constraint on the cardinalities of the particle and hole roots' sets:
\beq
0\, = \, \Oint{  \msc{C} }{}  \f{\dd \mu }{ 2\i\pi } \f{ \wh{\mf{a}}^{\, \prime}_{\mathbb{Y}}(\mu)  }{1 + \wh{\mf{a}}_{\mathbb{Y}}(\mu) } \, = \, 
|\wh{\mf{X}}| \, - \, |\wh{\mc{Y}}|  \, - \, s \;.
\enq
In particular, in the $s=0$ sector, $|\wh{\mf{X}}| \, =\, |\wh{\mc{Y}}|$.
Also, the introduction of $\wh{ \mathbb{Y} }$ allows us to make use of the below conventions:  
\beq
\sul{\la \in \wh{ \mathbb{Y} }  }{}   f(\la)   \; = \;  \sul{y \in \wh{\mc{Y}} }{}  f(y)       \, - \,  \sul{x \in \wh{\mf{X}} }{}  f(x) \qquad \e{and} \qquad 
\pl{\la \in\wh{ \mathbb{Y} }  }{}   f(\la)   \; = \;  \f{ \pl{y \in \wh{\mc{Y}} }{}  f(y)   }{ \pl{x \in \wh{\mf{X}} }{}  f(x)  } \;. 
\label{definition conventions somme et produits ensembles signes}
\enq

Following the standard procedure, one gets the non-linear integral equation satisfied by $ \wh{\mf{a}}_{\mathbb{Y}}(\xi) = \ex{ \wh{\mf{A}}_{\mathbb{Y}}(\xi) }$: 
\beq
\wh{\mf{A}}_{\mathbb{Y}}(\xi) \, = \,  -\f{h}{T} \, +\, \mf{w}_N(\xi)   - \; \i \pi s  \;   \; + \; \i \sul{ y \in \wh{\mathbb{Y}}  }{} \th(\xi-y)
\; + \; \Oint{  \msc{C}   }{} K(\xi-u) \cdot  \msc{L}\mathrm{n}\Big[ 1+  \ex{ \wh{\mf{A}}_{\mathbb{Y}} } \, \Big](u)  \cdot \dd u   
\label{ecriture eqn NLI forme primordiale}
\enq
with $\xi \in \mc{S}_{\zeta_{\e{m}}/2}$, the strip of width $\zeta_{\e{m}}=\e{min}(\zeta, \pi-\zeta)$ around $\R$. Above, for  $v\in \msc{C} $, one has
\beq
\msc{L}\mathrm{n} \Big[ 1+  \ex{ \wh{\mf{A}}_{\mathbb{Y}} } \, \Big](v) \, = \, \Int{\kappa}{v} \f{  \wh{\mf{A}}^{\, \prime}_{\mathbb{Y}}\!(u) }{ 1+  \ex{- \wh{\mf{A}}_{\mathbb{Y}}(u)} }  \cdot \dd u 
\; + \; \ln \Big[ 1+  \ex{ \wh{\mf{A}}_{\mathbb{Y}}(\kappa) } \, \Big] \;.
\enq
Here $\kappa$ is some point on $\msc{C} $ and the integral is taken, in the positive direction along $\msc{C} $, from $\kappa$ to $v$. The function $"\ln"$ appearing above corresponds to the principal branch of the logarithm 
extended to $\R^{-}$ with the convention $\e{arg}(z) \in \intfo{-\pi}{\pi}$. The functions $\theta$ and $K$ are respectively defined as
\begin{align}
   &\theta(\lambda) =\begin{cases}
      \displaystyle{ \i \ln\left(\frac{\sinh(\i \zeta+\lambda)}{\sinh(\i\zeta-\lambda)}\right)  }\quad 
                             &\text{for}\quad |\Im(\lambda)|<\zeta_m, \vspace{0.5mm}\\
      \displaystyle{ -\pi \, \mf{s}_2 +\i\ln\left(\frac{\sinh(\i \zeta+\lambda)}{\sinh(\lambda-\i\zeta)}\right) }
                             &\text{for}\quad \zeta_m< |\Im(\lambda)|<\pi/2,
     \end{cases}
     \label{def-theta}\\
    &K(\lambda)=\frac 1{2\pi} \theta'(\lambda)=\frac{\mf{s}_2}{2\pi\i}\,\big[\coth(\lambda-\i\zeta_m)-\coth(\lambda+\i\zeta_m)\big]
    =\frac{\sin(2\zeta)}{2\pi\,\sinh(\lambda-\i\zeta)\,\sinh(\lambda+\i\zeta)}\; ,
    \label{def-K}
\end{align}
whereas
\beq
\mf{w}_N(\xi) \; = \;  N \ln \bigg(  \f{ \sinh(  \xi - \tf{ \be }{N} ) \sinh(  \xi + \tf{ \be }{N} -\i\zeta) }{  \sinh(  \xi + \tf{ \be }{N} ) \sinh( \xi - \tf{ \be }{N} -\i\zeta )  }   \bigg)    \;.
\enq
One should note that the particle and hole roots forming $\wh{\mathbb{Y}}$ are subject to the subsidiary conditions 
\beq
\wh{\mf{a}}_{ \mathbb{Y} }(x) \, = \, -1 \quad \forall x \in \wh{\mf{X}} \qquad  \e{and} \qquad \wh{\mf{a}}_{ \mathbb{Y} }(y) \, = \, -1 \quad \forall y \in \wh{\mc{Y}}\, .  
\enq

In the remaining part of the paper, we shall use the following  convenient shorthand notation for an appropriate determination of the logarithm of $1+\wh{\mf{a}}_{\mathbb{Y}}$:
\beq
\label{notation log 1+a}
\wh{\msc{L}}_{\mathbb{Y}}(s) \, = \, \f{1}{2\i\pi} \msc{L}n\Big( 1+\wh{\mf{a}}_{\mathbb{Y}}\Big)(s) \;.
\enq

\subsubsection{The non-linear equation in the infinite Trotter number limit}
\label{sec-Trotter-limit}

To compute the infinite Trotter number limit, one \textit{assumes} that $\wh{\mf{A}}_{\mathbb{Y}} \limit{N}{+\infty} \mf{A}_{\mathbb{Y}}$ pointwise on $ \msc{C}$, 
and that all properties of the non-linear problem are preserved under this limit. Then,  one may readily deduce a non-linear integral equation satisfied by
$\mf{A}_{\mathbb{Y}}$. This procedure has been set into a rigorous setting, for $T$ large enough, in \cite{KozGohmannGoomaneeSuzukiRigorousApproachQTMForFreeEnergy}. 
Here we stick with a formal exposure and refer the interested reader to the mentioned work for more analytic details.

In the infinite Trotter number limit, one has 
\beq
 \mf{w}_N(\xi) \;  \limit{N}{+\infty}\;  - 2 \be \,\Big( \coth(\xi)-\coth(\xi-\i\zeta) \Big)= \f{2J \sin^2(\zeta) }{ T \sinh(\xi)\sinh(\xi-\i\zeta) } \;. 
\enq
In addition one \textit{assumes} the existence of the limit of the particle and hole roots in the below sense. First, one parameterises $\wh{\mf{X}}= \big\{ \wh{x}_a \big\}_1^{ | \wh{\mf{X}} | }$ and 
$\wh{\mc{Y}}= \big\{ \wh{y}_a \big\}_1^{ | \wh{\mc{Y}} | }$, so that, when $N \tend +\infty$
\beq
\label{limit-particle-hole-roots}
\wh{ x }_a\tend x_a + \i\f{\zeta}{2} \; , \;\; a=1,\dots,|\mf{X}| \qquad \e{and} \qquad  \wh{y}_a \tend y_a + \i\f{\zeta}{2} \; , \;\; a=1,\dots,|\mc{Y}| \;. 
\enq
All these assumptions result in the non-linear integral equation satisfied by the limit function on $\mc{S}_{\zeta_{\e{m}}/2}$, the strip of
width $\zeta_{\e{m}}$ centered around $\R$:
\beq
\mf{A}_{\mathbb{Y}}(\xi) \, = \, -\f{1}{T} e_{0}(\xi) \; - \; \i \pi s  \; 
+ \; \i \sul{ y \in \mathbb{Y} }{} \th(\xi-y)
\; + \; \Oint{ \msc{C}  }{} K(\xi-u) \cdot   \msc{L}\mathrm{n}\Big[ 1+ \ex{ \mf{A}_{\mathbb{Y}} } \Big](u) \cdot \dd u   \;.  
\label{ecriture NLIE Trotter infini}
\enq
Here, 
\beq
e_0(\xi) \, = \, h \, - \, \f{2J \sin^2(\zeta) }{ \sinh(\xi) \sinh(\xi-\i\zeta) } \;. 
\enq
Furthermore,   
\beq
\mathbb{Y} \, = \, \Big\{ \mc{Y} \, + \,   \i\tfrac{\zeta}{2} \Big\}    \ominus  \Big\{ \mf{X} \, + \,   \i\tfrac{\zeta}{2} \Big\}   \; , 
\label{premiere definition de Y kappa}
\enq
where 
\beq
\mf{X} \; = \; \big\{ x_a  \big\}_{a=1}^{ | \mf{X} | } \qquad \e{and} \qquad \mc{Y} \; = \; \big\{ y_a  \big\}_{a=1}^{ | \mc{Y} | } \;. 
\enq
The non-linear integral equation at infinite Trotter number is to be supplemented with the constraints
\beq
0\;=\; \Oint{ \msc{C}}{} \f{ \mf{A}^{\prime}_{\mathbb{Y}}(u) } { 1+ \ex{-\mf{A}_{\mathbb{Y}}(u)} }  \cdot \f{\dd u}{2\i\pi}  \; = \;|\mf{X}|  - |\mc{Y}| - s
\quad \e{and} \quad 
\left\{ \ba{cccc} \mf{a}_{ \mathbb{Y} }(x+\i\tfrac{\zeta}{2}) & = & -1 & \forall x \in \mf{X}  \\
  \mf{a}_{ \mathbb{Y} }(y+\i\tfrac{\zeta}{2}) & =  & -1 &  \forall y \in \mc{Y} \ea \right. \; ,
\label{ecriture condition subsidiaires solution NLIE spectre}
\enq
with $\mf{a}_{ \mathbb{Y} }(\xi)= \ex{\mf{A}_{\mathbb{Y}}(\xi)}$.
Similarly as in \eqref{notation log 1+a}, we shall also denote
\beq
\label{limite notation log 1+a}
\msc{L}_{\mathbb{Y}}(s) \, = \, \f{1}{2\i\pi} \msc{L}n\Big( 1+{\mf{a}}_{\mathbb{Y}}\Big)(s) \;.
\enq

\subsubsection{The Eigenvalues of the quantum transfer matrix}

One readily obtains the below representation for the Eigenvalues of the transfer matrix, labelled by the particle-hole roots' parameters $\wh{\mathbb{Y}}$:
\beq
\wh{\tau}_{\mathbb{Y}}(0) \, = \, \pl{y \in \wh{\mathbb{Y} }}{} \f{ \sinh(y-\i \zeta) }{ \sinh(y)} \cdot \bigg( \f{ \sinh(\tf{\be}{N}+\i\zeta) }{ \sinh(\i\zeta) }\bigg)^{2N}
\cdot \exp\Bigg\{ \f{h}{2T} \, - \, \Oint{ \msc{C} }{}  \f{ \i \sin(\zeta)  \msc{L}\mathrm{n} \Big[ 1+  \ex{ \wh{\mf{A}}_{\mathbb{Y}} } \, \Big](u) }{ \sinh(u-\i\zeta)\sinh(u) }  \f{\dd u}{2\i\pi}  \Bigg\} \;. 
\enq
The Trotter limit can easily be taken on the level of this representation, leading to $\wh{\tau}_{\mathbb{Y}}(0)\tend \tau_{\mathbb{Y}}(0)$ where 
\beq
\tau_{\mathbb{Y}}(0) \, = \, \pl{y \in \mc{Y}\ominus \mf{X} }{} \f{ \sinh(y-\i \tf{\zeta}{2}) }{ \sinh(y+\i \tf{\zeta}{2})}  
\cdot \exp\Bigg\{ \f{h}{2T}\, - \, \f{ 2J  }{T} \cos(\zeta) \, - \, \Oint{ \msc{C}-\i\tfrac{\zeta}{2} }{}  \f{ \i \sin(\zeta) \msc{L}\mathrm{n} \Big[ 1+  \ex{ \mf{A}_{\mathbb{Y}} } \, \Big](u+\i\zeta/2)  }{ \sinh(u-\i \tf{\zeta}{2})\sinh(u+\i \tf{\zeta}{2}) }  
\f{\dd u}{2\i\pi}  \Bigg\} \;. 
\enq

 \subsection{The boundary factor}
 
 We now explain how to represent the boundary factor \eqref{boundary factor} in terms of appropriate contour integrals involving the corresponding counting function $\wh{\mf{a}}_{\mathbb{Y}}$ solution of \eqref{ecriture eqn NLI forme primordiale}.

  It was established in \cite{KozPozsgaySurfaceFreeEnergyBoundaryXXZ} that, for any set of Bethe roots $\{\mu_a\}_1^N$ at magnetic field $h^{\prime}$ solving \eqref{ecriture BAE QTM} and  labelled by the particle-hole roots' parameters $\wh{\mathbb{Y}}$, the partition function \eqref{ecriture Tsuchiya determinant} involved in the expression of the boundary factor  \eqref{boundary factor} can be represented in the following form:
\beq
\mc{Z}_N\Big( \{-\tfrac{\be}{N}\}_1^N ; \{ \mu_a \}_1^N;\xi_- \Big) \,  =  \, \mc{P}(\{ \mu_a \}_1^N ) \cdot \ex{ \wh{\msc{F}}_{\mathbb{Y}} } \cdot 
\bigg( \f{  1 + \wh{\mf{a}}_{\mathbb{Y}}(0) }{ 1-\wh{\mf{a}}_{\mathbb{Y}}(0) }\bigg)^{\f{1}{4}}    \; .  
 \label{lemme factorisation det ac Fn}
\enq
The latter involves the counting function $\wh{\mf{a}}_{\mathbb{Y}}$ that has been defined in \eqref{definition fct auxiliaire a}, 
a pure product function $\mc{P}$ which reads
\beq
\label{def-fct-P}
 \mc{P}(\{ \mu_a \}_1^N ) \, = \, \pl{a=1}{N} G(\mu_a) \cdot \pl{a<b}{N} f(\mu_a,\mu_b) 
\enq
with
\begin{align}
&f(\la,\mu)\, = \, \f{ \sinh(\la+\mu-\i\zeta) }{ \sinh(\la+\mu) }, \label{def-f}\\
&G(\mu) \, = \,  \f{ \sinh(2\tf{\be}{N} ) \sinh(\mu + \xi_-) }{ \sinh(2\mu) } \Big\{ \sinh(\mu-\tfrac{\be}{N}) \sinh(\i\zeta +\mu+\tfrac{\be}{N} ) \Big\}^{N}  \cdot  \Big\{ 1+\wh{\mf{a}}_{\mathbb{Y}}(-\mu) \Big\}^{\f{1}{2}} \; ,
\label{def-G}
\end{align}
and a contribution $\wh{\msc{F}}_{\mathbb{Y}}$ which is given by a series of multiple contour integrals:
\bem
 \wh{\msc{F}}_{\mathbb{Y}} \, = \, 
\sul{ k = 0  }{+\infty}   \Oint{ \msc{C}^{(1)}\supset\dots \supset \msc{C}^{(2k+1)} }{}
\hspace{-2mm} 
\sul{ n=k }{ +\infty } \f{ \Big[ \wh{\mf{r}}_{\mathbb{Y}} (\om_{2k+1}) \Big]^{n-k} }{ 2n + 1 }  \; \cdot  \; 
 \pl{p=1}{2k+1}\wh{U}_{\mathbb{Y}}(\om_p,\om_{p+1}) \; \f{ \dd^{2k+1} \om }{ (2\i\pi)^{2k+1} }  \\
\; -\; \sul{  k = 1   }{+\infty}   \Oint{   \msc{C}^{(1)}\supset\dots \supset \msc{C}^{(2k)} }{}
\sul{n=k}{+\infty} \f{ \Big[ \wh{\mf{r}}_{\mathbb{Y}}(\om_{2k}) \Big]^{n-k} }{2n} \; \cdot \;  \pl{p=1}{2k} \wh{U}_{\mathbb{Y}}(\om_p,\om_{p+1}) \; \f{ \dd^{2k} \om }{ (2\i\pi)^{2k} } \; . 
\label{ecriture msc hat F a trotter fini}
\end{multline}
In \eqref{ecriture msc hat F a trotter fini}, the integration contours are encased contours such that, for any $p$, $\msc{C}^{(1)}\supset\dots \supset \msc{C}^{(p)}$, and that $ \msc{C}^{(k)}$, 
$k=1,\dots,p$, enlaces the roots $\mu_1, \dots, \mu_N$ but not the ones that are shifted by $\pm \i\zeta$. 
The integration variables in \eqref{ecriture msc hat F a trotter fini} satisfy to the convention $\om_{2k+2}\equiv \om_1$ while the integrands are built up from the function 
\begin{align}\label{r-fct}
\wh{\mf{r}}_{\mathbb{Y}}(\om) 
&= \ex{ -\f{2h^{\prime} }{T} } \pl{a=1}{N} \f{ \sinh(\mu_a+\om+\i\zeta) \sinh(\mu_a-\om+\i\zeta)  }
				{ \sinh(\mu_a+\om-\i\zeta)\sinh(\mu_a-\om - \i\zeta)  }  \; ,
%
%
\end{align}
as well as the kernel 
\begin{align}\label{U-kernel}
\wh{U}_{\mathbb{Y}}(\om,\om^{\prime})  &= \f{ - \ex{ -\f{h^{\prime} }{T} } \sinh(2\om^{\prime}-\i\zeta) }
			{   \sinh(\om + \om^{\prime} )\sinh(\om-\om^{\prime} - \i\zeta)}  
\pl{a=1}{N} \f{ \sinh( \om^{\prime}+\mu_a  ) \sinh(\om^{\prime} - \mu_a - \i\zeta)  }
				{  \sinh(\om^{\prime}-\mu_a) \sinh( \om^{\prime} +\mu_a -\i\zeta) }    \; .
%
%
%
%
%
\end{align}
%
%
%

In \cite{KozPozsgaySurfaceFreeEnergyBoundaryXXZ}, the representation \eqref{lemme factorisation det ac Fn}, for the particular case of the Bethe roots $\{\lambda_a\}_1^N$ describing the dominant Eigenstate of the quantum transfer matrix, was reformulated in a smooth way for the consideration of the Trotter limit, in terms of contour integrals involving the function $\wh{\mf{a}}_{\mathbb{\emptyset}}$.
We now explain how to similarly rewrite, in terms of the function $\wh{\mf{a}}_{\mathbb{Y}}$, the partition function \eqref{lemme factorisation det ac Fn} for the more general case of a set of Bethe roots $\{\mu_a\}_1^N$ describing some sub-dominant state of the quantum transfer matrix labelled by the particle-hole roots' parameters $\wh{\mathbb{Y}}$.
For the purpose of the calculations to come, it is useful to introduce an auxiliary set of roots $\{\nu_a\}_1^N$
which corresponds to the collection of all the roots of $1+\wh{\mf{a}}_{\mathbb{Y}}$ located inside of $\msc{C}$:
\beq
\{\mu_a\}_1^N\, = \, \{\nu_a\}_1^{N-n}\cup \{\, \wh{y}_a\}_1^n  \qquad \e{and} \qquad \{\nu_a\}_1^N\, = \, \{\nu_a\}_1^{N-n}\cup \{\wh{x}_a\}_1^n
\label{definition des racines nu et decomposition des racines mu}
\enq
 where we chose the parameterisation of the sets $\wh{\mc{Y}}=\{\wh{y}_a\}_1^n$ and $\wh{\mf{X}}=\{\wh{x}_a\}_1^n$.

For a sub-dominant state of the quantum transfer matrix as described above, one can factorise the product function $\mc{P}$ \eqref{def-fct-P} in the form
\beq
\mc{P}\big( \{\mu_a\}_1^N \big) \; = \; \mc{P}\big( \{\nu_a\}_1^N \big) \cdot \mc{P}_{\e{bk}}\Big( \{\nu_a\}_1^N; \wh{\mathbb{Y}} \Big) \cdot \mc{P}_{\e{loc}}\Big( \wh{\mathbb{Y}} \Big) \;, 
\label{decomposition preliminaire de mathcalP}
\enq
in which, using the product convention \eqref{definition conventions somme et produits ensembles signes}, we have defined 
\begin{align}
&\mc{P}_{\e{bk}}\Big( \{\nu_a\}_1^N; \wh{\mathbb{Y}}  \Big) \; = \; \pl{a=1}{N} \pl{z \in \wh{\mathbb{Y}} }{}f(\nu_a,z) \;, \\
&\mc{P}_{\e{loc}}\Big( \wh{\mathbb{Y}} \Big) \; = \; \f{ \pl{a<b}{n} \Big\{ f(\, \wh{y}_a,\wh{y}_b) \, f(\, \wh{x}_a,\wh{x}_b) \Big\}   }{ \pl{a,b=1}{n} f(\, \wh{x}_a,\wh{y}_b)   }
\pl{a=1}{n} \bigg\{ \f{  G(\wh{y}_a) }{  G(\wh{x}_a) }  f(\,\wh{x}_a,\wh{x}_a)  \bigg\} \;. 
\end{align}
This decomposition  follows readily from the product identity 
\beq
 \pl{a<b}{N} f(\mu_a,\mu_b)  \; = \;  \pl{a<b}{N}  f(\nu_a,\nu_b) \cdot  \pl{a=1}{N}  \pl{b=1}{n} \f{ f(\nu_a,\wh{y}_b)  }{ f(\nu_a,\wh{x}_b)  } \cdot 
\f{ \pl{a<b}{n}\Big\{ f(\, \wh{y}_a, \wh{y}_b) f(\, \wh{x}_a,\wh{x}_b) \Big\}   }{ \pl{a,b=1}{n} f(\, \wh{x}_a,\wh{y}_b)   }
\cdot \pl{a=1}{n} f(\, \wh{x}_a,\wh{x}_a) \;. 
\enq
The rewriting of $\mc{P}$ in terms of contour integrals is based on the following lemma concerning the rewriting of products over the Bethe roots $\nu_a$, $1\le a\le N$:

 \begin{lemme}
\label{Lemme reecriture plusieurs produits} 
  
Let $\{\mu_a\}_1^N$ be a solution of the Bethe Ansatz equations \eqref{ecriture BAE QTM}, $\wh{\mf{a}}_{\mathbb{Y}}$ its associated counting function  and let the roots $\{\nu_a\}_1^N$ be defined as in \eqref{definition des racines nu et decomposition des racines mu}. 
Then,  for $z$ an arbitrary parameter, we have
\begin{align}
&\pl{a=1}{N} \sinh(\nu_a+z)  \, = \,   \Big\{ \sinh(z- \tfrac{\be}{N} )\Big\}^{N}   \cdot  \Big\{ 1+\wh{\mf{a}}_{\mathbb{Y}}(-z)
\Big\}^{ \bs{1}_{ \msc{D}}(-z) } 
\cdot \exp\bigg\{-  \Oint{ \msc{C} }{} \dd s \,\wh{\msc{L}}_{\mathbb{Y}}(s)  \coth\big( s + z  \big)  \bigg\} \, ,
\label{ecriture produit general avec un z}
\\
&\pl{a=1}{N} \sinh(2\nu_a+z)  \, = \,   \Big\{ \sinh(z- 2\tfrac{\be}{N} )\Big\}^{N}   \cdot  
\Big\{ 1+\wh{\mf{a}}_{\mathbb{Y}}(-\frac{z}{2}) 
\Big\}^{ \bs{1}_{ \msc{D}}(-\frac z2) } 
\cdot \exp\bigg\{-  2 \Oint{ \msc{C} }{} \dd s \,\wh{\msc{L}}_{\mathbb{Y}}(s)  \coth\big( 2 s + z  \big)  \bigg\} \, ,
\label{prod-sinh2nu-z}
\end{align}
as well as
\begin{multline}\label{id-double-prod-z}
  \pl{a,b=1}{N} \sinh(\nu_a+\nu_b+z) \, = \,
  \Big\{ \sinh(z- \tfrac{2\be}{N} )\Big\}^{N^2} \cdot 
  \Big\{ 1+\wh{\mf{a}}_{\mathbb{Y}}(-z+  \tfrac{\be}{N}) \Big\}^{N\cdot \bs{1}_{ \msc{D}}(-z+  \tfrac{\be}{N}) } \cdot
  \prod_{a=1}^N \Big\{ 1+\wh{\mf{a}}_{\mathbb{Y}}(-z-\nu_a) \Big\}^{\bs{1}_{ \msc{D}}(-z-\nu_a) } \\
  \times
  \exp\bigg\{ \Oint{\msc{C} }{} \dd s  \Oint{\msc{C}^{\prime} \subset \msc{C} }{} \dd s^{\prime} \, \coth^{\prime}(s+s^{\prime}+z) \,\wh{\msc{L}}_{\mathbb{Y}}(s) \, \wh{\msc{L}}_{\mathbb{Y}}(s^{\prime})  
 \, - \, 2N \Oint{ \msc{C} }{} \dd s \,\wh{\msc{L}}_{\mathbb{Y}}(s)\, \coth(s-\tfrac{\be}{N}+z) \bigg\} 
 \, .
\end{multline}
Here $\bs{1}_{ \msc{D}}(\omega) $ is equal to $1$ when $\omega\in\msc{D}$ mod $\i\pi \mathbb{Z}$ and to $0$ otherwise.
In \eqref{id-double-prod-z} the contour $\msc{C}^{\prime} \subset \msc{C}$ has to be chosen so that it encircles the points $\nu_a$, $1\le a \le N$, as well as the neighbourhood of the origin, but not the poles at $s'=-s-z\,$     for $s\in\msc{C}$.
\end{lemme}

 \Proof  
  
So as to estimate the products \eqref{ecriture produit general avec un z}, one introduces the function
\beq
\label{SN1}
\mc{S}^{(1)}_N(\om) \, = \, \sul{a=1}{N}   \ln \sinh(\nu_a+\om + z) \, .
\enq
Its derivative can be written as
\beq
 \Big( \mc{S}^{(1)}_N \Big)^{\prime}(\om) \, = \,  \Oint{ \msc{C}  }{} \f{\dd s }{2\i\pi} \coth(s + \om + z) \f{ \wh{\mf{a}}_{\mathbb{Y}}^{ \, \prime}(s) }{ 1 \, + \,  \wh{\mf{a}}_{\mathbb{Y}}(s)  }
\, + \, N \coth( \om -\tfrac{\be}{N} +  z) \, - \, \bs{1}_{ \msc{D} }\big( - \om - z \big)\f{ \wh{\mf{a}}_{\mathbb{Y}}^{ \, \prime}( - \om - z) }{ 1 \, + \,  \wh{\mf{a}}_{\mathbb{Y}}( - \om - z)  }\, .
\enq
Upon taking the ante-derivative and then the exponent, this yields
\beq
  \ex{ \mc{S}^{(1)}_N (\om) } \, = \,  \Big\{ \sinh( \om -\tfrac{\be}{N} +  z) \Big\}^{N} \cdot 
  \Big[  1 \, + \,  \wh{\mf{a}}_{\mathbb{Y}}( - \om - z) \Big]^{ \bs{1}_{ \msc{D} }\big( - \om - z \big) }  \ex{ C^{(1)}(\om)} 
\cdot \exp\bigg\{ - \Oint{ \msc{C}  }{} \dd s \coth(s + \om + z) \wh{\msc{L}}_{\mathbb{Y}}(s) \bigg\}\;. 
\label{ecriture expression avec constante pour exp de S1}
\enq
There, $C^{(1)}(\om)$ is $\i\pi$-periodic and constant on $-\msc{D}-z$ and on $\big\{ \la \; : \; |\Im(\la)| \leq \tf{\pi}{2}\big\} \setminus \big\{ -\ov{\msc{D}}-z \big\} $. 
However, observe that, by construction, the function $ \om \mapsto \ex{ \mc{S}^{(1)}_N (\om) }$ is continuous across $-\msc{C}-z$. Also, it is easy to see by using the Sokhotsky-Plemejl formulae that the function
\beq
\om \mapsto   \Big[  1 \, + \,  \wh{\mf{a}}_{\mathbb{Y}}( - \om - z) \Big]^{ \bs{1}_{ \msc{D} }\big( - \om - z \big) } \cdot   \exp\bigg\{ - \Oint{ \msc{C}  }{} \dd s \coth(s + \om + z)\, \wh{\msc{L}}_{\mathbb{Y}}(s) \bigg\}
\enq
is also continuous across $-\msc{C}-z$. The two formulae can thus match only if $C^{(1)}(\om)$ is continuous across $-\msc{C}-z$. Thus $C^{(1)}(\om)=C^{(1)}$ does not depend on $\om$. 
One can then fix its value by comparing the $\om\tend +\infty$ behaviour of the two sides of \eqref{ecriture expression avec constante pour exp de S1}. 
On the one hand, it is direct to infer from the finite product representation issued from \eqref{SN1} that 
\beq
   \ex{ \mc{S}^{(1)}_N (\om) } \, \widesim{ \om \tend + \infty} \,  \f{1}{2^N} \, \ex{  N (\om +z) }\cdot \pl{a=1}{N} \ex{ \nu_a }  \;. 
\enq
On the other hand, one has that 
\beq
 -\Oint{ \msc{C}  }{} \dd s \,\coth(s + \om + z) \,\wh{\msc{L}}_{\mathbb{Y}}(s) \, 
 \widesim{ \om \tend + \infty} \,   -\Oint{ \msc{C}  }{} \dd s   \,\wh{\msc{L}}_{\mathbb{Y}}(s)  
\, =\,   \Oint{ \msc{C}  }{} \f{ \dd s }{2\i\pi} \,  s  \, \f{ \wh{\mf{a}}_{\mathbb{Y}}^{ \, \prime}(s) }{ 1 \, + \,  \wh{\mf{a}}_{\mathbb{Y}}(s)  } 
\, = \, \sul{a=1}{N} \nu_a \; +\; \be \;,
\enq
which leads to
\bem
\big\{ \sinh( \om -\tfrac{\be}{N} +  z) \big\}^{N} \cdot 
  \Big[  1 \, + \,  \wh{\mf{a}}_{\mathbb{Y}}( - \om - z) \Big]^{ \bs{1}_{ \msc{D} }\big( - \om - z \big) }  \ex{ C^{(1)}} 
\cdot \exp\bigg\{ - \Oint{ \msc{C}  }{} \dd s \coth(s + \om + z)\, \wh{\msc{L}}_{\mathbb{Y}}(s) \bigg\} \\
 \, \widesim{ \om \tend + \infty} \, \f{1}{2^N}\, \ex{  N (\om +z) + C^{(1)}}\cdot \pl{a=1}{N} \ex{ \nu_a } \;. 
\end{multline}
 By comparing the form of these asymptotics, one gets that $C^{(1)}=0$\;. The identity \eqref{ecriture produit general avec un z} hence follows by setting $\om=0$.
 
 The product \eqref{prod-sinh2nu-z} can be computed similarly.

Let us now establish \eqref{id-double-prod-z}. Using \eqref{ecriture produit general avec un z} we can write
\begin{multline}\label{double-pd-1}
     \pl{a,b=1}{N} \sinh(\nu_a+\nu_b+z) \, 
     = \,
     \pl{a=1}N \Bigg\{ \Big[ \sinh(\nu_a+ z- \tfrac{\be}{N} )\Big]^{N}  \,  
     \Big[ 1+\wh{\mf{a}}_{\mathbb{Y}}(-\nu_a-z) \Big]^{ \bs{1}_{ \msc{D}}(-\nu_a-z) } 
     \\
     \times
     \exp\bigg[-   \Oint{ \msc{C} }{} \dd s \,\wh{\msc{L}}_{\mathbb{Y}}(s)  \coth\big(  s + \nu_a + z  \big)  \bigg] \Bigg\}\; .
\end{multline}
Using again  \eqref{ecriture produit general avec un z} to compute the first product in \eqref{double-pd-1},
\begin{multline}
   \pl{a=1}N  \Big[ \sinh(\nu_a+ z- \tfrac{\be}{N} )\Big]^{N}  \, = \, \Big\{ \sinh(z- 2\tfrac{\be}{N} )\Big\}^{N^2}   \cdot  \Big\{ 1+\wh{\mf{a}}_{\mathbb{Y}}(-z+\tfrac{\beta}{N})\Big\}^{ N\cdot \bs{1}_{ \msc{D}}(-z+\f{\beta}{N}) } 
  \\
 \times \exp\bigg\{- N  \Oint{ \msc{C} }{} \dd s \,\wh{\msc{L}}_{\mathbb{Y}}(s)  \coth\big( s + z- \tfrac{\be}{N}  \big)  \bigg\} \, ,
\end{multline}
and observing that 
\begin{align}
 & -   \sul{a=1}{N}\Oint{ \msc{C} }{} \dd s \,\wh{\msc{L}}_{\mathbb{Y}}(s)  \coth\big(  s + \nu_a + z  \big)  \nonumber\\
%
%
\; &\hspace{2cm}= \; -\Oint{ \msc{C} }{} \dd s  \hspace{-2mm}  \Oint{ \msc{C}^{\prime} \subset \msc{C} }{} \hspace{-2mm} \f{ \dd s^{\prime} }{ 2\i\pi } 
 \coth\big( s + s^{\prime} +z \big) \, \wh{\msc{L}}_{\mathbb{Y}}(s) \, \f{ \wh{\mf{a}}_{\mathbb{Y}}^{ \, \prime}(s^{\prime}) }{ 1 \, + \,  \wh{\mf{a}}_{\mathbb{Y}}(s^{\prime})  }   
 \; - \; N \Oint{ \msc{C} }{} \dd s \,\wh{\msc{L}}_{\mathbb{Y}}(s) \, \coth\Big( s +z - \tfrac{\be}{N} \Big)  \nonumber\\
 \; &\hspace{2cm}= \; 
 \Oint{ \msc{C} }{} \dd s  \hspace{-2mm}  \Oint{ \msc{C}^{\prime} \subset \msc{C} }{} \hspace{-2mm}  \dd s^{\prime}   
 \coth^{\prime}\big( s + s^{\prime} +z \big) \, \wh{\msc{L}}_{\mathbb{Y}}(s) \, \wh{\msc{L}}_{\mathbb{Y}}(s^{\prime}) 
 \; - \; N \Oint{ \msc{C} }{} \dd s \,\wh{\msc{L}}_{\mathbb{Y}}(s) \, \coth\Big( s +z - \tfrac{\be}{N} \Big) \;,
\end{align}
we obtain \eqref{id-double-prod-z}.
\qed 
 
This lemma allows us to formulate the following proposition:
 
\begin{prop}

The partition function \eqref{ecriture Tsuchiya determinant} involved in the expression of the boundary factor  \eqref{boundary factor} can be represented as in \eqref{lemme factorisation det ac Fn}.

In this expression, the product function $\mc{P}$ can be rewritten as
\begin{multline}\label{expr-P}
\mc{P}\big( \{\mu_a\}_1^N \big) \; = \;  \mf{p}_N \cdot  \mc{K}^{(1)}_N \big( \wh{\mathbb{Y}} \big) \cdot 
\f{ \big[ 1+\wh{\mf{a}}_{\mathbb{Y}}(-\xi_-) \big]^{ \bs{1}_{\msc{D}}(-\xi_-) }  }{  \big[ 1 + \wh{\mf{a}}_{\mathbb{Y}}(0) \big]^{\f{1}{2}} }\cdot 
\exp\Bigg\{ \Oint{ \msc{C} }{} \dd s  \Oint{ \msc{C}^{\prime}\subset \msc{C} }{} \dd s^{\prime}\, \phi^{(+)}_2(s,s^{\prime})\, \wh{\msc{L}}_{\mathbb{Y}}(s)\, \wh{\msc{L}}_{\mathbb{Y}}(s^{\prime}) \\
\, + \, \Oint{ \msc{C} }{} \dd s \,\phi_1(s)\, \wh{\msc{L}}_{\mathbb{Y}}(s)  
\, - \, N \Oint{ \msc{C} }{} \dd s \, \wh{\msc{L}}_{\mathbb{Y}}(s) \, \Dp{s} \ln \sinh(s,\i\zeta +\tfrac{\be}{N} ) \Bigg\} \; ,
\end{multline}
in which
\beq
\mf{p}_N \; = \;  \Bigg\{ \f{\sinh(\i\zeta+2\tf{\be}{N}) }{ \sinh(2\tf{\be}{N}) }\Bigg\}^{\f{N(N-1)}{2}}\cdot 
\Big\{\sinh(\i\zeta)\, \sinh\Big( 2\tfrac{\be}{N} \Big)  \Big\}^{N^2} \cdot  \Big\{ \sinh\Big( \xi_- - \tfrac{\be}{N} \Big) \Big\}^{N} \;. 
\enq
and 
\bem
\label{K_N^1}
\mc{K}^{(1)}_N\big( \wh{\mathbb{Y}} \big) \, = \, \f{ \pl{a<b}{n}f(\wh{y}_a,\wh{y}_b) f(\wh{x}_a,\wh{x}_b)  }{ \pl{a,b=1}{n} f(\wh{x}_a,\wh{y}_b)   } \cdot 
\pl{a=1}{n} \bigg\{  \f{ \sinh(\wh{y}_a + \xi_-) }{ \sinh(\wh{x}_a + \xi_-) } \cdot \f{ \sinh(2 \, \wh{x}_a -\i\zeta) }{ \sinh(2 \, \wh{y}_a) } \cdot
 \Big\{ 1+\wh{\mf{a}}_{\mathbb{Y}}(-\wh{x}_a) \Big\}^{\f{1}{2}}  \cdot  \Big\{ 1+\wh{\mf{a}}_{\mathbb{Y}}(-\wh{y}_a) \Big\}^{\f{1}{2}} \bigg\}  \\
\times    \pl{a=1}{n} \bigg\{ \f{ \sinh(\i\zeta+\tf{\be}{N},\wh{y}_a) }{ \sinh(\i\zeta+\tf{\be}{N},\wh{x}_a) }   \bigg\}^N \cdot 
\pl{z \in   \wh{\mathbb{Y}}  }{} \exp\bigg\{ - \Oint{ \msc{C} }{}   \dd  s  \, \wh{\msc{L}}_{\mathbb{Y}}(s) \,  \Dp{s} \ln f(s,z) \bigg\} \;. 
\end{multline}
Note that $\msc{C}$ is taken here such that it also satisfies the property that if $x \in \wh{\mf{X}}$ then also $-x \in \e{Int}(\msc{C})$.
Also, we have introduced the functions 
\begin{align}
&\phi^{(+)}_2(s,s^{\prime}) \; = \; \f{1}{2} \Big[ \coth^{\prime}(s+s^{\prime}-\i\zeta)  \, - \, \coth^{\prime}(s+s^{\prime}) \Big] \;,
\label{phi2+}\\ 
&\phi_1(s) \, = \, \coth(2s-\i\zeta) \, + \, \coth(2s) \, - \,\coth(s+\xi_-) \;,
\label{phi1}
\end{align}
and the shortcut notation
\begin{equation}\label{double-sinh}
 \sinh(s,s')\, =\, \sinh(s+s')\sinh(s-s')\;.
\end{equation}
The contours are chosen in such a way that the poles of the integral involving $\phi_1$ at $\i\tf{\zeta}{2}$ and $\i\tf{(\zeta-\pi)}{2}$ are located outside of $\msc{C}$. 
Likewise, in the double integral involving $\phi^{(+)}_2$, the contours are chosen in such a way that the poles at $s+s^{\prime}-\i\zeta$ are always located outside from the 
contours $\msc{C}$ or $\msc{C}^{\prime}$. 

The contribution $\wh{\msc{F}}_{\mathbb{Y}}$ is given as in \eqref{ecriture msc hat F a trotter fini}, in which the kernel $\wh{U}_{\mathbb{Y}}$ is given by
\bem\label{kernelU-new}
\wh{U}_{\mathbb{Y}}(\om,\om^{\prime}) = \f{ - \ex{-\f{h^{\prime}}{T}} \s{2\om^{\prime}-\i\zeta} }{ \s{\om+\om^{\prime}} \s{\om-\om^{\prime}-\i\zeta}}
\cdot \bigg[ \f{ \s{\om^{\prime}- \tf{\be }{N}} \s{\om^{\prime}+ \tf{\be }{N}-\i\zeta} }
	{ \s{\om^{\prime}+ \tf{\be }{N}} \s{\om^{\prime} - \tf{\be }{N}-\i\zeta} }  \bigg]^N \\ 
\times
\exp\Bigg\{  - \Oint{ \msc{C}_U }{}  \dd  s  \, \wh{\msc{L}}_{\mathbb{Y}}(s) \, 
\Big[ \coth(\om^{\prime} +s) \, + \, \coth( \om^{\prime} -s )
 \, -  \, \coth(\om^{\prime}+s -\i\zeta) \, - \, \coth(\om^{\prime} -s- \i\zeta)  \Big] \Bigg\}	\;, 
\end{multline}
and the function $\wh{\mf{r}}_{\mathbb{Y}}$ by
\bem\label{fct-r-new}
\wh{\mf{r}}_{\mathbb{Y}}(\om) = \ex{ -\f{2h^{\prime}}{T} } \bigg[ \f{ \s{\om- \tf{\be }{N}+\i\zeta} \s{\om+ \tf{\be }{N}-\i\zeta} }
	{ \s{\om+ \tf{\be }{N} +\i\zeta } \s{\om - \tf{\be }{N}-\i\zeta} }  \bigg]^N \\ 
\times
\exp\Bigg\{  - \Oint{ \msc{C}_U }{}  \dd  s  \, \wh{\msc{L}}_{\mathbb{Y}}(s) \, 
\Big[ \coth(s +\om +\i\zeta) \, + \, \coth(\om -s +\i\zeta)
 \, -  \, \coth(\om +s -\i\zeta) \, - \, \coth(\om -s -\i\zeta)  \Big]  \Bigg\}	\;. 
\end{multline}
In the above expressions, the contour $\msc{C}_U $ encircles the points $\{\mu_a\}_1^N$ as well as the origin, and is such that, given any $\om\in\msc{C}^{(p)}$, where $\msc{C}^{(p)}$ refers to any of the encasted contours introduced in 
\eqref{ecriture msc hat F a trotter fini}, the points $\pm\om,\ \pm(\om+\i\zeta),\, \pm(\om-\i\zeta)$ are not surrounded by $\msc{C}_U $.
%
%
%
%
%
%
%
%

\end{prop}

\Proof  

The formulas \eqref{kernelU-new} and \eqref{fct-r-new} are direct rewritings of \eqref{U-kernel} and \eqref{r-fct}. Let us therefore prove \eqref{expr-P}.

Using \eqref{id-double-prod-z} with $z=0$ and $z=-\i\zeta$, together with the decomposition
\beq
\pl{a<b}{N} \sinh(\nu_a+\nu_b+z) \, = \, \f{ \bigg\{ \pl{a,b=1}{N} \sinh(\nu_a+\nu_b+z)  \bigg\}^{\f{1}{2}}  }{ \bigg\{ \pl{a=1}{N} \sinh(2\nu_a+z)  \bigg\}^{\f{1}{2}}  } \;,
\enq
and observing that $\wh{\mf{a}}_{\mathbb{Y}}(\tfrac{\beta}{N})=0$,
we obtain that
\bem
\label{identity-prod1}
\pl{a<b}{N} \sinh(\nu_a+\nu_b) \, = \, \Big\{ \sinh(-2\tfrac{\be}{N})\Big\}^{\f{N^2}{2}} \cdot \pl{a=1}{N} \Bigg\{ \f{  1+\wh{\mf{a}}_{\mathbb{Y}}(-\nu_a)  }{ \sinh(2\nu_a) }  \Bigg\}^{\f{1}{2}}
 \\
\times 
\exp\bigg\{ \f{1}{2} \Oint{\msc{C}  }{} \dd \om  \Oint{\msc{C}^{\prime} \subset \msc{C}  }{} \dd \om^{\prime} \coth^{\prime}(\om+\om^{\prime})\, \wh{\msc{L}}_{\mathbb{Y}}(\om) \, \wh{\msc{L}}_{\mathbb{Y}}(\om^{\prime}) 
\,-\, N \Oint{ \msc{C} }{} \dd \om \,\wh{\msc{L}}_{\mathbb{Y}}(\om) \coth(\om-\tfrac{\be}{N})  \bigg\}\; ,
\end{multline}
and 
\bem
\label{identity-prod2}
\pl{a<b}{N} \sinh(\nu_a+\nu_b-\i\zeta) \, = \, \Big\{ \sinh(-\i\zeta -2\tfrac{\be}{N})\Big\}^{\f{N^2}{2}} \cdot
\pl{a=1}{N} \Bigg\{ \f{ 1 }{ \sinh(2\nu_a-\i\zeta) }  \Bigg\}^{\f{1}{2}}
  \\
\times 
\exp\bigg\{ \f{1}{2} \Oint{\msc{C} }{} \dd \om  \Oint{\msc{C}^{\prime} \subset \msc{C} }{} \dd \om^{\prime} \coth^{\prime}(\om+\om^{\prime}-\i\zeta) \,
\wh{\msc{L}}_{\mathbb{Y}}(\om) \, \wh{\msc{L}}_{\mathbb{Y}}(\om^{\prime})  
\, -\, N \Oint{ \msc{C} }{} \dd \om \,\wh{\msc{L}}_{\mathbb{Y}}(\om) \coth(\om-\tfrac{\be}{N}-\i\zeta) \bigg\}  \, .
\end{multline}
Here the integration contour $\msc{C}^{\prime} \subset \msc{C}$ is chosen such that the poles at $\om'=-\om$ and at $\om^{\prime}=-\om+\i\zeta$ all lie outside of $\msc{C}^{\prime}$ whereas those at $\om=-\om'+\i\zeta$ lie outside of $\msc{C}$. 
Using also \eqref{prod-sinh2nu-z}, we get
\beq
 \pl{a=1}{N} \bigg\{ \f{ \sinh(2\nu_a) }{  \sinh(2\nu_a-\i\zeta) }  \bigg\}^{\f{1}{2}}  \, = \,   \bigg\{  \f{   \sinh( 2\tfrac{\be}{N})  }{ \sinh(\i\zeta+2\tfrac{\be}{N}) } \bigg\}^{\f{N}{2}} 
\big[ 1 + \wh{\mf{a}}_{\mathbb{Y}}(0) \big]^{\f{1}{2}}    
\exp\bigg\{  \Oint{ \msc{C} }{} \dd s \wh{\msc{L}}_{\mathbb{Y}}(s)  \Big[ \coth(2s-\i\zeta) \, - \, \coth(2s)   \Big]    \bigg\}.
\enq
It follows that
\begin{align}
\pl{a<b}{N} f(\nu_a,\nu_b) \, 
%
%
&= \, \bigg\{  \f{  \sinh(\i\zeta + 2\tfrac{\be}{N})  }{   \sinh(2\tfrac{\be}{N})  }\bigg\}^{\f{N(N-1)}{2}}\cdot
      \frac{\big[1 + \wh{\mf{a}}_{\mathbb{Y}}(0)\big]^{\f{1}{2}} }{\pl{a=1}N \big[ 1+\wh{\mf{a}}_{\mathbb{Y}}(-\nu_a) \big]^{\f{1}{2}} } \cdot
      \exp\bigg\{  \Oint{ \msc{C} }{} \dd s \wh{\msc{L}}_{\mathbb{Y}}(s)  \Big[  \coth(2s-\i\zeta)\, -\, \coth(2s)    \Big]    \bigg\}
\nonumber\\
&\hspace{2cm}\times 
\exp\bigg\{ N \Oint{ \msc{C} }{} \dd \om\, \wh{\msc{L}}_{\mathbb{Y}}(\om) \Big[ \coth(\om-\tfrac{\be}{N} )  -  \coth(\om-\tfrac{\be}{N}-\i\zeta) \Big] \bigg\} 
\nonumber\\
&\hspace{2cm}\times 
\exp\bigg\{  \Oint{\msc{C} }{} \dd \om  \Oint{\msc{C}^{\prime} \subset \msc{C} }{} \!\dd \om^{\prime} \,\phi^{(+)}_2(\om,\om^{\prime})\,
\wh{\msc{L}}_{\mathbb{Y}}(\om) \wh{\msc{L}}_{\mathbb{Y}}(\om^{\prime})  \bigg\} \; ,
\end{align}
in which $\phi^{(+)}_2$ is given by \eqref{phi2+}.

Moreover, it also follows from \eqref{ecriture produit general avec un z}  that
\begin{align}
&\pl{a=1}{N} \sinh(\nu_a+\xi_-) \, = \,  \sinh^N(\xi_- - \tfrac{\be}{N})  \cdot \big[1+\wh{\mf{a}}_{\mathbb{Y}}(-\xi_-)\big]^{ \bs{1}_{ \msc{D}}(-\xi_-) }
\cdot \exp\bigg\{- \Oint{ \msc{C} }{} \dd s\, \wh{\msc{L}}_{\mathbb{Y}}(s)\, \coth(s+\xi_-) \bigg\} \, ,
\label{ecriture identite pour produit avec xi-}
\\
&\pl{a=1}{N} \sinh^{N}(\nu_a-\tfrac{\be}{N})  \, = \,   \sinh^{N^2}  (-2 \tfrac{\be}{N} )
\cdot \exp\bigg\{- N \Oint{ \msc{C} }{} \dd s \, \wh{\msc{L}}_{\mathbb{Y}}(s) \, \coth\big( s - \tfrac{\be}{N} \big)  \bigg\} \, ,
\label{ecriture produit avec  moins beta sur N}
\\
&\pl{a=1}{N} \sinh^{N}(\i\zeta+\nu_a+\tfrac{\be}{N})\, = \, \sinh^{N^2} (\i\zeta) 
\cdot \exp\bigg\{- N  \Oint{ \msc{C} }{} \dd s\, \wh{\msc{L}}_{\mathbb{Y}}(s)\, \coth(s+\i\zeta+\tfrac{\beta}{N}) \bigg\} \, ,
\end{align}
and from \eqref{prod-sinh2nu-z} that
\begin{equation}
 \prod_{a=1}^N\sinh(2\nu_a)\, =\, \sinh^N(- 2\tfrac{\be}{N} )   \cdot  \big[ 1+\wh{\mf{a}}_{\mathbb{Y}}(0) \big]
\cdot \exp\bigg\{-  2 \Oint{ \msc{C} }{} \dd s \,\wh{\msc{L}}_{\mathbb{Y}}(s)  \coth\big( 2 s   \big)  \bigg\} \, ,
\end{equation}
so that
\begin{multline}
  \prod_{a=1}^N G(\nu_a) \, =\, 
   \frac{ \pl{a=1}{N} \big[  1 + \wh{\mf{a}}_{\mathbb{Y}}(-\nu_a) \big]^{\f{1}{2}}  \cdot \big[1+\wh{\mf{a}}_{\mathbb{Y}}(-\xi_-)\big]^{\bs{1}_{\msc{D}}(-\xi_-) } }{1+\wh{\mf{a}}_{\mathbb{Y}}(0)}
\cdot \Big\{    \sinh(\i \zeta  )  \,  \sinh\Big( 2 \tfrac{\be}{N} \Big)   \Big\}^{N^2 }  \sinh^N(\xi_- -\tfrac{\be}{N}) \\
\times \exp\bigg\{ \Oint{ \msc{C} }{} \dd \om \,\wh{\msc{L}}_{\mathbb{Y}}(\om)\, \Big[ 2 \coth(2\om)-\coth(\om+\xi_-) - N  \coth(\om-\tfrac{\be}{N} ) - N  \coth(\om+\tfrac{\be}{N}+\i\zeta) \Big]  \bigg\}   \;. 
\end{multline}

Finally, 
\begin{align}
\mc{P}_{\e{bk}}\Big( \{\nu_a\}_1^N; \wh{\mathbb{Y}} \Big) \; 
&= \; \prod_{a=1}^N \left\{ \pl{y \in \wh{\mc{Y}} }{}  \frac{\sinh(\nu_a+y-\i\zeta)}{\sinh(\nu_a+y)}  \; \pl{x \in \wh{\mf{X}} }{}   \frac{\sinh(\nu_a+x)}{\sinh(\nu_a+x-\i\zeta)}  \right\} 
\nonumber\\
&= \;  \pl{z \in \wh{\mathbb{Y}} }{} \left\{ \Big[ f(- \tfrac{\be}{N},z )\Big]^{N} 
         \exp\bigg[-  \Oint{ \msc{C}  }{} \dd s \, \wh{\msc{L}}_{\mathbb{Y}}(s) \, \Dp{s} \ln f(s,z) \bigg] \right\} 
  \cdot \pl{x \in \wh{\mf{X}} }{} \big[ 1+\wh{\mf{a}}_{\mathbb{Y}}(-x)\big]
  \; ,
\end{align}
in which we have used \eqref{ecriture produit general avec un z} with $z=y,y-\i\zeta$ for $y \in \wh{\mc{Y}}$, $-y$ and $-y+\i\zeta$
being outside of $\msc{D}$ mod $\i\pi \mathbb{Z}$, and with $z=x,x-\i\zeta$ for $x \in \wh{\mf{X}}$, $-x\in\msc{D}$ and $-x+\i\zeta\notin\msc{D}$ mod $\i\pi \mathbb{Z}$.

It then remains to put all these partial results together to obtain \eqref{expr-P}. 
\qed

\smallskip


As a result, we can formulate the following rewriting of the boundary factor \eqref{boundary factor} at finite Trotter number:

\begin{prop}\label{prop-boundary-factor}
Let $\{\la_a(h)\}_1^{N}$ stand for the Bethe roots describing the dominant state of the quantum transfer matrix $\op{t}_{\mf{q};h}$ at magnetic field $h$, and $\{\mu_a(h^{\prime})\}_1^{N}$ be a set of Bethe roots for a sub-dominant state of the quantum transfer matrix $\op{t}_{\mf{q};h'}$ at magnetic field $h'$, characterised by a set \eqref{set-def} $\wh{\mathbb{Y}}$.

Then, the boundary factor \eqref{boundary factor} admits the following representation:
\begin{multline}\label{Boundary-factor-finite-Trotter}
%
\mc{F}_{\mc{B}}\big( \{ \mu_a(h^{\prime}) \}_1^N ; \{\lambda_a(h)\}_1^N; \xi_- \big) 
\; = \;  \ex{ \wh{\msc{F}}_{\mathbb{Y}}- \wh{\msc{F}}_{\emptyset} } \cdot 
\Bigg[ \f{  1 - \wh{\mf{a}}_{\emptyset}(0)^2 }{ 1-\wh{\mf{a}}_{\mathbb{Y}}(0)^2 }\Bigg]^{\f{1}{4}}    \cdot
\Bigg[ \f{  1 +\wh{\mf{a}}_{\mathbb{Y}}(-\xi_-) }{ 1+\wh{\mf{a}}_{\emptyset}(-\xi_-) }\Bigg]^{\bs{1}_{\msc{D}}(-\xi_-) } \!\! \cdot \,
\mc{K}^{(1)}_N \big( \wh{\mathbb{Y}} \big) \\
\times
\exp\Bigg\{ \Oint{ \msc{C} }{} \dd s \!\!  \Oint{ \msc{C}^{\prime}\subset \msc{C} }{} \!\! \dd s^{\prime}\, \phi^{(+)}_2(s,s^{\prime})\, \left[ \wh{\msc{L}}_{\mathbb{Y}}(s)\, \wh{\msc{L}}_{\mathbb{Y}}(s^{\prime})\, -\,\wh{\msc{L}}_{\emptyset}(s)\, \wh{\msc{L}}_{\emptyset}(s^{\prime})\right]  \\
\, + \, \Oint{ \msc{C} }{} \dd s \,\phi_1(s)\, \left[ \wh{\msc{L}}_{\mathbb{Y}}(s)  \, -\, \wh{\msc{L}}_{\emptyset}(s)\right]
\, - \, N \Oint{ \msc{C} }{} \dd s \, \left[ \wh{\msc{L}}_{\mathbb{Y}}(s)\, -\, \wh{\msc{L}}_{\emptyset}(s)\right] \, \Dp{s} \ln \sinh(s,\i\zeta +\tfrac{\be}{N} ) \Bigg\} \; .
\end{multline}
In this expression, $\wh{\mf{a}}_{\mathbb{Y}}\equiv\wh{\mf{a}}_{\mathbb{Y};h'}$ and $\wh{\mf{a}}_{\emptyset}\equiv\wh{\mf{a}}_{\emptyset;h}$ stand for the respective counting functions of $\{\mu_a(h^{\prime})\}_1^{N}$ and $\{\la_a(h)\}_1^{N}$, whereas $\wh{\msc{L}}_{\mathbb{Y}}$ and $\wh{\msc{L}}_{\emptyset}$ are defined from these counting functions as in \eqref{notation log 1+a}.
The contribution $\wh{\msc{F}}_{\mathbb{Y}}$ is given as in \eqref{ecriture msc hat F a trotter fini}, in which the kernel $\wh{U}_{\mathbb{Y}}\equiv\wh{U}_{\mathbb{Y};h'}$ and the function $\wh{\mf{r}}_{\mathbb{Y}}\equiv\wh{\mf{r}}_{\mathbb{Y};h'}$ are given respectively by \eqref{kernelU-new} and \eqref{fct-r-new} with $h$ replaced by $h'$, the contribution $\wh{\msc{F}}_{\emptyset}$ being defined similarly in terms of $h$ and $\wh{\msc{L}}_{\emptyset}$.
The functions $\phi^{(+)}_2$ and $\phi_1$ are defined in \eqref{phi2+} and \eqref{phi1}.
The factor $\mc{K}^{(1)}_N\big( \wh{\mathbb{Y}} \big)$ is expressed in terms of the particle-hole configuration of the set of Bethe roots $\{\mu_a(h^{\prime})\}_1^{N}$ as in \eqref{K_N^1}.
Finally, $\msc{C}$ has the property that if $x \in \wh{\mf{X}}$ then $-x \in \e{Int}(\msc{C})$.

\end{prop}

\subsection{The matrix element}

We now repeat the analysis relatively to the relevant matrix element to the problem. For that purpose, we need to recall a few know facts. 

The scalar product between two on-shell Bethe vectors at different magnetic fields and parameterised by roots $\{\la_a(h)\}_1^{N}$
and $\{\mu_a(h^{\prime})\}_1^{N}$ have been shown \cite{KozKitMailSlaTerXXZsgZsgZAsymptotics} to admit the determinant representation
\bem
\label{det-repr-SP}
\Big( \bs{\Psi}(\{ \la_a\}_1^N)  ,   \bs{\Psi}(\{ \mu_a \}_1^N) \Big) \; = \; \Big(  \bs{\Psi}(\{ \mu_a\}_1^N) ,  \bs{\Psi}(\{ \la_a\}_1^N)  \Big) \;= \;
  \pl{a=1}{N}\bigg\{  a_{h^{\prime}}(\la_a) d_{ h^{\prime} }(\mu_a)    \pl{b=1}{N}\f{ \sinh(\mu_b-\la_a-\i\zeta) }{  \sinh(\la_a-\mu_b) }     \bigg\}  \\
\times \pl{b=1}{N} \bigg\{ \varkappa \f{ V_+(\la_b) }{ V_-(\la_b) } \, - \, 1  \bigg\} \cdot \f{ 1\, - \,  \varkappa  }{ V_+^{-1}(\th) \, - \, \varkappa  V_-^{-1}(\th)     } 
\cdot \det_{N}\Big[ \de_{jk}  \; + \; U_{jk}^{(\la)}(\th) \Big] \;,
\end{multline}
for $\varkappa=\ex{ \f{ h - h^{\prime} }{ T } }$ and $\theta$ an arbitrary complex number. Here we have used the following notations:
\begin{align}
 &a_{h}(\la) \; = \; \ex{ \f{h}{2T}}  \bigg\{ \f{ \sinh(\la-\tfrac{\be}{N}-\i\zeta) \sinh(\la+\tfrac{\be}{N} ) }{ \sinh^2(-\i\zeta) }  \bigg\}^{N}\; ,\\
 &d_{h}(\la) \; = \; \ex{ -\f{h}{2T}}  \bigg\{ \f{ \sinh(\la+\tfrac{\be}{N}+\i\zeta) \sinh(\la - \tfrac{\be}{N} )  }{ \sinh^2(-\i\zeta) } \bigg\}^{N}\; ,\\
&V_{\pm}(\om) \; = \; \pl{b=1}{N} \f{ \s{\la_b-\om \mp \i\zeta} }{   \s{\mu_b-\om \mp \i\zeta} } \;, 
\end{align}
and  
\beq
U_{jk}^{(\la)}(\th)  \,  =  \,    \f{ \pl{a=1}{N} \sinh\big(\mu_a-\la_j\big) }{  \pl{ \substack{ a=1 \\ \not= j } }{N} \sinh(\la_a-\la_j) }
\cdot \f{ K_{\varkappa}(\la_j-\la_k)  \; - \; K_{\varkappa}(\th-\la_k) }{ V_+^{-1}(\la_j) \; - \; \varkappa V_-^{-1}(\la_j) }  ,
\enq
in which
\beq
\label{K-kappa}
K_{\varkappa}( \om ) \; = \; \coth(\la+\i\zeta) \; - \; \varkappa \coth (\la-\i\zeta) \;. 
\enq
Note that, if  $\{\la_a(h)\}_1^{N}$ stand for the Bethe roots describing the dominant state of the quantum transfer matrix $\op{t}_{\mf{q};h}$ and $\{\mu_a(h^{\prime})\}_1^{N}$ a sub-dominant state of the quantum transfer matrix $\op{t}_{\mf{q};h'}$ characterised by a set \eqref{set-def} $\wh{\mathbb{Y}}$, we can write
\beq
\label{ratioV-a}
\varkappa \f{ V_+(\om) }{ V_-(\om) } \; = \;  \f{ \wh{\mf{a}}_{\mathbb{Y}}(\om) }{ \wh{\mf{a}}_{\emptyset}(\om)  }\; ,
\enq
in terms of the corresponding counting functions $\wh{\mf{a}}_{\mathbb{Y}}\equiv\wh{\mf{a}}_{\mathbb{Y};h'}$ and $\wh{\mf{a}}_{\emptyset}\equiv\wh{\mf{a}}_{\emptyset;h}$.
Similarly as in \cite{KozKitMailSlaTerXXZsgZsgZAsymptotics}, we can also rewrite the finite size determinant in \eqref{det-repr-SP} as a Fredholm determinant of an integral operator acting on a the contour surrounding the points $\{\la_a(h)\}_1^{N}$. For latter convenience, we choose this contour  $\Ga(\msc{C})$ to surround also the contour $\msc{C}$. 
Hence \eqref{det-repr-SP} can be rewritten as 
\bem\label{formule-SP}
\Big( \bs{\Psi}(\{ \la_a\}_1^N)  ,   \bs{\Psi}(\{ \mu_a \}_1^N) \Big) \; = \; (-1)^N  
  \pl{k=1}{N}\bigg\{  a_{h^{\prime}}(\la_k) d_{ h^{\prime} }(\mu_k)    \pl{b=1}{N}\f{ \sinh(\mu_b-\la_k-\i\zeta) }{  \sinh(\la_k-\mu_b) }     \bigg\}  \\
\times \pl{b=1}{N} \Big( \wh{\mf{a}}_{\mathbb{Y}}(\la_b)  \, + \, 1  \Big) \cdot \f{ (1\, - \,  \varkappa)V_+(\th)   }{ 1\, - \,\tfrac{ \wh{\mf{a}}_{\mathbb{Y}}(\th) }{ \wh{\mf{a}}_{\emptyset}(\th)  }    } 
\cdot \det_{\Ga(\msc{C}) }\Big[ \op{id}  \; + \; \wh{\op{U}}_{\th}^{(\la)} \Big] \;. 
\end{multline}
There, we have set
\begin{align}
\wh{\op{U}}_{\th}^{(\la)}(\om,\om^{\prime})  \,  &=  \,    -\frac{1}{2\pi \i}\pl{a=1}{N}  \f{ \sinh\big(\mu_a-\om\big) }{  \sinh(\la_a-\om) }
\cdot \f{ K_{\varkappa}(\om-\om^{\prime})  \; - \; K_{\varkappa}(\th-\om^{\prime}) }{ V_+^{-1}(\om) \; - \; \kappa V_-^{-1}(\om) }  \nonumber\\
&=\,  -\frac{V_+(\om)}{2\pi \i}\pl{a=1}{N}  \f{ \sinh\big(\mu_a-\om\big) }{  \sinh(\la_a-\om) }
\cdot \f{ K_{\varkappa}(\om-\om^{\prime})  \; - \; K_{\varkappa}(\th-\om^{\prime}) }{ 1\, - \,\tfrac{ \wh{\mf{a}}_{\mathbb{Y}}(\om) }{ \wh{\mf{a}}_{\emptyset}(\om)  } }  \, .
\end{align}

Finally, one recalls the "norm" formula for Bethe vectors. Given any solution $\{\mu_a\}_1^N$ to the Bethe equation at non-zero
magnetic field $h'$, and a contour $\msc{C}_{\mathbb{Y}}$ which surrounds  the points $\{\mu_a(h')\}_1^{N}$ as well as a neighbourhood of the origin, one has that 
\beq\label{formule-norme}
\Big( \bs{\Psi}(\{ \mu_a\}_1^N) ,   \bs{\Psi}(\{ \mu_a\}_1^N) \Big)  \; = \; 
\pl{ s=1 }{ N } \bigg\{ \f{  \wh{\mf{a}}_{\mathbb{Y}}^{ \, \prime}(\mu_s) }{  \wh{\mf{a}}_{\mathbb{Y}}(\mu_s) } \cdot a_{h^{\prime}}(\mu_s)d_{h^{\prime}}(\mu_s) \bigg\}
\cdot \f{ \pl{a,b=1}{N} \sinh(\mu_a-\mu_b - \i\zeta)  }{  \pl{ \substack{ a \not= b} }{ N } \sinh(\mu_a-\mu_b) }  
\, \cdot \, \det_{ \msc{C}_{\mathbb{Y}} } \Big[ \op{id} \; + \; \wh{\ov{K}}_{\mathbb{Y}} \Big] \;. 
\enq
In the above, $\wh{\ov{K}}_{\mathbb{Y}}$ is an integral operator on $L^2( \msc{C}_{\mathbb{Y}} )$ with integral kernel
\beq
\wh{\ov{K}}_{\mathbb{Y}}(\om,\om^{\prime}) \; = \; -\f{   K(\om - \om^{\prime} ) }{  1 \, + \,  \wh{\mf{a}}_{\mathbb{Y}}^{-1}(\om^{\prime}) }
\label{definition noyau integral hat ov K Y}
\enq
Although this is not directly necessary here for our purpose of computing the infinite Trotter number limit, it is also possible to re-express the Fredholm determinant in \eqref{formule-norme} in terms of an integral operator acting on the contour $\msc{C}$ instead of $\msc{C}_\mathbb{Y}$. We refer to Appendix~C of \cite{KozGohmannKarbachSuzukiFiniteTDynamicalCorrFcts} for details about such a reformulation.

Before stating the main result of the section, we first need to establish a technical lemma.

\begin{lemme}
\label{Lemme rep int pour rations PS input periodique}

It holds
\bem
 \f{  \pl{ \substack{ a \not= b} }{ N } \sinh(\nu_a-\nu_b) }{   \pl{  a,  b=1 }{ N } \sinh(\nu_a-\la_b)    } \cdot 
 \pl{b=1}{N} \Bigg\{   \f{   \wh{\mf{a}}_{\mathbb{Y}}(\la_b)  \, + \, 1  }{   \ln^{\prime} \wh{\mf{a}}_{\mathbb{Y}}(\nu_b)  }  \Bigg\}  \pl{a=1}{N}\Bigg\{ \f{ \sinh(\la_a+\tf{\be}{N}) }{ \sinh(\nu_a+\tf{\be}{N}) } \Bigg\}^N  \\
 \, = \,  \exp\Bigg\{   \Oint{ \msc{C} }{} \dd s  \Oint{ \msc{C}^{\prime}\subset \msc{C} }{} \dd s^{\prime} \coth^{\prime}(s-s^{\prime})
\wh{\msc{L}}_{\mathbb{Y}}(s)\, \Big[\wh{\msc{L}}_{\emptyset}(s^{\prime}) \, - \,  \wh{\msc{L}}_{\mathbb{Y}}(s^{\prime})  \Big] \Bigg\} \;. 
\label{produit diagonal des nua contre nua et la b}
\end{multline}
Also, for $|\eps|$ small enough and generic, it holds 
\bem
 \pl{z \in \wh{\mathbb{Y}} }{} \pl{a=1}{N}  \bigg\{ \f{ \sinh(\la_a-z-\eps) }{\sinh(\nu_a-z+\eps) \sinh(\nu_a-z-\eps) } \bigg\}  \; = \;
 \pl{z \in\wh{\mathbb{Y}} }{} \bigg\{ \f{1 }{\sinh(\eps-z-\tfrac{\be}{N})  }  \bigg\}^{N}
\pl{ x \in \wh{\mf{X}}  }{}  \bigg\{ \f{ \big[1+ \wh{\mf{a}}_{\mathbb{Y}}(x-\eps) \big]  \big[1+ \wh{\mf{a}}_{\mathbb{Y}}(x+\eps) \big]  }{   \big[1+ \wh{\mf{a}}_{\emptyset}(x+\eps) \big] }   \bigg\} \\
\times \pl{z \in \wh{\mathbb{Y}} }{} \exp\Bigg\{    \Oint{  \msc{C} }{} \dd s \coth(s-z+\eps) \wh{\msc{L}}_{\mathbb{Y}}(s)\, - \,  \coth(s-z - \eps) \Big[\wh{\msc{L}}_{\emptyset}(s) \, - \,  \wh{\msc{L}}_{\mathbb{Y}}(s)  \Big] \Bigg\}  \;. 
\label{ecriture explicite produit sur les z}
\end{multline}

\end{lemme}

\Proof

By virtue of \eqref{ecriture produit general avec un z}, it is easy to see that, for $\epsilon$ small enough,
\begin{align}
\pl{  a,  b=1 }{ N }  \f{    \sinh(\nu_a-\la_b + \eps)    }{ \sinh(\nu_a-\nu_b + \eps) } \, &= \, \pl{a=1}{N}\Bigg\{ \f{ \sinh(\la_a-\eps +\tf{\be}{N}) }{ \sinh(\nu_a-\eps+\tf{\be}{N}) } \Bigg\}^N 
\cdot 
 \pl{b=1}{N} \Bigg\{   \f{  1  \, + \,  \wh{\mf{a}}_{\mathbb{Y}}(\la_b-\eps)    }{  1  \, + \, \wh{\mf{a}}_{\mathbb{Y}}(\nu_b-\eps)  }  \Bigg\}   \\
 \, & \times  \,  \exp\Bigg\{  - \Oint{ \msc{C} }{} \dd s  \Oint{ \msc{C}^{\prime}\subset \msc{C} }{} \dd s^{\prime} \coth^{\prime}(s-s^{\prime}+\eps)
\wh{\msc{L}}_{\mathbb{Y}}(s)\, \Big[\wh{\msc{L}}_{\emptyset}(s^{\prime}) \, - \,  \wh{\msc{L}}_{\mathbb{Y}}(s^{\prime})  \Big] \Bigg\}\, .
\end{align}
Then, the $\eps\tend 0^+$ limit yields \eqref{produit diagonal des nua contre nua et la b}.  The representation \eqref{ecriture explicite produit sur les z}
follows from  \eqref{ecriture produit general avec un z} through even more direct handlings. \qed

\smallskip

This lemma enables us to formulate the following rewriting of the ratio of matrix elements \eqref{ratio elements matrices} that is used for the expression \eqref{expr-QN} of the one-point function \eqref{expr-1pt-fct}:

\begin{prop}
\label{prop-matrix-el-finite-Trotter}
Let $\{\la_a(h)\}_1^{N}$ stand for the Bethe roots describing the dominant state of the quantum transfer matrix $\op{t}_{\mf{q};h}$ at magnetic field $h$, and $\{\mu_a(h^{\prime})\}_1^{N}$ be a set of Bethe roots for a sub-dominant state of the quantum transfer matrix $\op{t}_{\mf{q};h'}$ at magnetic field $h'$, characterised by a set \eqref{set-def} $\wh{\mathbb{Y}}$.

Then, the ratio of matrix elements \eqref{det-repr-SP} and \eqref{formule-norme} admits the following representation:
%
%
%
\bem
 \f{ \Big( \bs{\Psi}(\{ \la_a\}_1^N)  ,   \bs{\Psi}(\{ \mu_a(h^{\prime}) \}_1^N) \Big) }
 { \Big( \bs{\Psi}(\{ \mu_a(h^{\prime}) \}_1^N) ,   \bs{\Psi}(\{ \mu_a(h^{\prime}) \}_1^N) \Big)  }  \; = \;
\mc{K}^{(2)}_N\big( \wh{\mathbb{Y}} \big) 
\cdot 
\exp\Bigg\{   \Oint{ \msc{C} }{} \dd s \!\! \Oint{ \msc{C}^{\prime}\subset \msc{C} }{} \!\! \dd s^{\prime}\, \phi^{(-)}_2(s,s^{\prime})\,
\wh{\msc{L}}_{\mathbb{Y}}(s)\, \Big[\wh{\msc{L}}_{\mathbb{Y}}(s^{\prime}) \, - \,  \wh{\msc{L}}_{\emptyset}(s^{\prime})  \Big]  \\
%
%
%
\; + \; N \Oint{ \msc{C} }{} \dd s \, \Big[\wh{\msc{L}}_{\mathbb{Y}}(s) \, - \,  \wh{\msc{L}}_{\emptyset}(s)  \Big] \, \Dp{s} \ln \sinh(s,\i\zeta +\tf{\be}{N} ) \Bigg\} 
\cdot 
\f{ (1\, - \,  \kappa)\, V_+(\th)   }{ 1\, - \,\tfrac{ \wh{\mf{a}}_{\mathbb{Y}}(\th) }{ \wh{\mf{a}}_{\emptyset}(\th)  }    }  \cdot 
\f{ \det_{\Ga(\msc{C}) }\Big[ \op{id}  \; + \; \wh{\op{U}}_{\th}^{(\la)} \Big] }{   \det_{ \msc{C}_{\mathbb{Y}} } \Big[  \op{id}  \; + \; \wh{\ov{K}}_{\mathbb{Y}} \Big]  }\, ,
\end{multline}
where the factor $\mc{K}^{(2)}_N\big( \wh{\mathbb{Y}} \big)$ is expressed in terms of the particle-hole
configuration of the set of Bethe roots $\{\mu_a(h^{\prime})\}_1^{N}$ as
\begin{multline}\label{K_N^2}
   \mc{K}^{(2)}_N\big( \wh{\mathbb{Y}} \big) \; = \;
   \pl{z \in \wh{\mathbb{Y}} }{} \Big\{  \sinh\big(z,\i\zeta+\tfrac{\be}{N} \big) \Big\}^{-N} \cdot
\f{ \pl{ \substack{z\not=z^{\prime} \\ z, z^{\prime}\in  \wh{\mathbb{Y}} } }{}  \sinh(z-z^{\prime}) }{   \pl{z, z^{\prime}\in \wh{\mathbb{Y}}  }{}  \sinh(z-z^{\prime}-\i\zeta)  } \cdot
\pl{ x \in \wh{\mf{X}} }{} \bigg\{  \f{   \wh{\mf{a}}_{\emptyset}(x)  \, + \, 1  }{-  \ln^{\prime} \wh{\mf{a}}_{\mathbb{Y}}(x)  }   \bigg\}  \cdot 
\pl{ y \in \wh{\mc{Y}} }{} \bigg\{  \f{   1  }{  \ln^{\prime} \wh{\mf{a}}_{\mathbb{Y}}(y)  }   \bigg\}   \\ 
\times\,   \pl{z \in \wh{\mathbb{Y}} }{} \exp\Bigg\{    \Oint{ \msc{C} }{} \dd s \, \bigg[\wt{\psi}_1(s,z)\, \wh{\msc{L}}_{\mathbb{Y}}(s)\, + \, \wt{\psi}_2(s,z)\,\Big[\wh{\msc{L}}_{\emptyset}(s) \, - \,  \wh{\msc{L}}_{\mathbb{Y}}(s)  \Big] \bigg]\Bigg\}   \, . 
\end{multline}
In these expressions, $\wh{\mf{a}}_{\mathbb{Y}}\equiv\wh{\mf{a}}_{\mathbb{Y};h'}$ and $\wh{\mf{a}}_{\emptyset}\equiv\wh{\mf{a}}_{\emptyset;h}$ stand for the respective counting functions of $\{\mu_a(h^{\prime})\}_1^{N}$ and $\{\la_a(h)\}_1^{N}$, whereas $\wh{\msc{L}}_{\mathbb{Y}}$ and $\wh{\msc{L}}_{\emptyset}$ are defined from these counting functions as in \eqref{notation log 1+a}.
We have also defined $\kappa=\ex{ \f{ h - h^{\prime} }{ T } }$, used the shortcut notation \eqref{double-sinh} and introduced the functions 
\begin{align}
&\phi^{(-)}_2(s,s^{\prime}) \; = \;  \coth^{\prime}(s-s^{\prime}-\i\zeta)  \, - \, \coth^{\prime}(s-s^{\prime}) \, ,\\
&\wt{\psi}_1(s,z) \, = \, \coth(s-z-\i\zeta)  \, - \,\coth(s-z) \, ,\\
&\wt{\psi}_2(s,z) \, = \, \coth(s-z)  \, - \,\coth(s-z+\i\zeta) \;.  
\end{align}
$\theta$ is here an arbitrary complex number such that $\th+\i\zeta\notin\msc{D}$, so that the coefficient $V_+(\th)$
can be rewritten as
\begin{equation}
V_+(\th)\,=\, \pl{ z \in \wh{\mathbb{Y}} }{ } \bigg\{ \f{1}{ \sinh(z-\th- \i\zeta) } \bigg\} \cdot  
%
\exp\Bigg\{ -  \Oint{ \msc{C} }{} \dd s \, \coth(s-\th- \i\zeta) \,\Big[\wh{\msc{L}}_{\emptyset}(s) \, - \,   \wh{\msc{L}}_{\mathbb{Y}}(s)  \Big]  \Bigg\}  \, .
\end{equation}
Finally, $\wh{\ov{K}}_{\mathbb{Y}}$ is an integral operator on $L^2( \msc{C}_{\mathbb{Y}} )$, where  $ \msc{C}_{\mathbb{Y}}$ surrounds the points $\{\mu_a(h^{\prime})\}_1^{N}$ and the origin, with integral kernel given by \eqref{definition noyau integral hat ov K Y}, whereas  $ \wh{\op{U}}_{\th}^{(\la)}$ is an integral operator on $L^2\Big( \Ga(\msc{C}) \Big)$, where $\Ga(\msc{C})$ surrounds the contour $\msc{C}$,  with integral kernel given as
\bem
 \wh{\op{U}}_{\th}^{(\la)}\big( \om , \om^{\prime}\big) \; = \; -\frac{1}{2\pi\i}\,
 \pl{z\in \wh{\mathbb{Y}} }{} \bigg\{ \f{ \sinh(z-\om) }{ \sinh(z-\om-\i\zeta)  } \bigg\} \cdot 
\exp\Bigg\{    \Oint{ \msc{C} }{} \dd s \, \Big[ \coth(s-\om-\i\zeta) \, - \, \coth(s-\om) \Big] \cdot \Big[\wh{\msc{L}}_{\mathbb{Y} }(s) \, - \,  \wh{\msc{L}}_{\emptyset}(s)  \Big] \Bigg\}  \\
\times  \f{ K_{\kappa}(\om-\om^{\prime})  \; - \; K_{\kappa}(\th-\om^{\prime}) }{  1\, - \,\tfrac{ \wh{\mf{a}}_{\mathbb{Y}}(\om) }{ \wh{\mf{a}}_{\emptyset}(\om)  }   } \;. 
\label{definition noyau integral hat U th lambda}
\end{multline}

 \end{prop}

\Proof 

It follows from the norm and scalar product formulae \eqref{formule-norme} and \eqref{formule-SP} that the  normalised twisted scalar product can be written as
\bem
 \f{ \Big( \bs{\Psi}(\{ \la_a\}_1^N)  ,   \bs{\Psi}(\{ \mu_a(h^{\prime}) \}_1^N) \Big) }
              { \Big( \bs{\Psi}(\{ \mu_a(h^{\prime})\}_1^N) ,   \bs{\Psi}(\{ \mu_a(h^{\prime})\}_1^N) \Big)  }\; = \;
\pl{b=1}{N} \Bigg\{   \f{   \wh{\mf{a}}_{\mathbb{Y}}(\la_b)  \, + \, 1  }{   \ln^{\prime} \wh{\mf{a}}_{\mathbb{Y}}(\mu_b)  }   \cdot \f{  a_{h^{\prime}}(\la_b)  }{  a_{h^{\prime}}(\mu_b)  }
\cdot \pl{a=1}{N} \f{  \sinh(\mu_a-\la_b - \i\zeta)   }{  \sinh(\mu_a-\mu_b - \i\zeta)    }    \Bigg\} \\
\times   \f{  \pl{ \substack{ a \not= b} }{ N } \sinh(\mu_a-\mu_b) }{   \pl{  a,  b=1 }{ N } \sinh(\mu_a-\la_b)    }
 \cdot  \f{ (1\, - \,  \varkappa) V_+(\th)   }{ 1\, - \,\tfrac{ \wh{\mf{a}}_{\mathbb{Y}}(\th) }{ \wh{\mf{a}}_{\emptyset}(\th)  }    } 
\cdot   \f{ \det_{ \Gamma(\msc{C}) }\Big[ \op{id}  \; + \; \wh{\op{U}}_{\th}^{(\la)} \Big] }{   \det_{ \msc{C}_\mathbb{Y} } \Big[ \e{id} \; + \; \wh{\ov{K}}_{\mathbb{Y}} \Big]  }\;. 
\end{multline}

Further, for any arbitrary $\epsilon$, one readily infers the product decomposition
\bem
\pl{  a,  b=1 }{ N }  \f{    \sinh(\mu_a-\la_b+\eps) }{ \sinh(\mu_a-\mu_b+\eps)    }  \; = \; 
\pl{b=1}{N}  \pl{a=1}{N} \bigg\{ \f{    \sinh(\nu_a-\la_b+\eps) }{ \sinh(\nu_a-\nu_b+\eps)    }    \bigg\}  \\
\times \pl{z \in \wh{\mathbb{Y}} }{} \pl{a=1}{N} \bigg\{ \f{  \sinh(z-\la_a + \eps)  }{ \sinh(\nu_a - z + \eps)  \sinh(z-\nu_a + \eps)   }  \bigg\} 
\cdot \pl{z, z^{\prime} \in \wh{\mathbb{Y}} }{} \bigg\{  \f{ 1  }{   \sinh(z - z^{\prime} + \eps)   }  \bigg\} \;. 
\end{multline} 
Then, it remains to invoke Lemma \ref{Lemme rep int pour rations PS input periodique} and take the $\eps \tend 0$ limit so as to obtain
\bem
 \f{  \pl{ \substack{ a \not= b} }{ N } \sinh(\mu_a-\mu_b) }{   \pl{  a,  b=1 }{ N } \sinh(\mu_a-\la_b)    } \cdot 
 \pl{b=1}{N} \Bigg\{   \f{   \wh{\mf{a}}_{\mathbb{Y}}(\la_b)  \, + \, 1  }{   \ln^{\prime} \wh{\mf{a}}_{\mathbb{Y}}(\mu_b)  }  \Bigg\}  \pl{a=1}{N}\Bigg\{ \f{ \sinh(\la_a+\tf{\be}{N}) }{ \sinh(\mu_a+\tf{\be}{N}) } \Bigg\}^N \\ 
\, = \, \pl{x \in \wh{\mf{X}} }{} \bigg\{ \f{     1 +  \wh{\mf{a}}_{\emptyset}(x) }{  -  \ln^{\prime} \wh{\mf{a}}_{\mathbb{Y}}(x)    }  \bigg\} 
\cdot  \pl{y \in \wh{\mc{Y}} }{} \bigg\{ \f{  1  }{    \ln^{\prime} \wh{\mf{a}}_{\mathbb{Y}}(y)    }  \bigg\}
\cdot  \pl{z \not= z^{\prime} \in \wh{\mathbb{Y}} }{} \Big\{     \sinh(z - z^{\prime} )   \Big\}  \cdot 
\pl{z \in \wh{\mathbb{Y}} }{}\exp\Bigg\{   \Oint{ \msc{C} }{} \dd s \, \coth(s-z)\, \Big[\wh{\msc{L}}_{\emptyset}(s) \, - \, 2 \wh{\msc{L}}_{\mathbb{Y}}(s)  \Big] \Bigg\}  \\
 \times  \exp\Bigg\{   \Oint{ \msc{C} }{} \dd s  \Oint{ \msc{C}^{\prime}\subset \msc{C} }{} \dd s^{\prime}\, \coth^{\prime}(s-s^{\prime})\,
\wh{\msc{L}}_{\mathbb{Y}}(s)\, \Big[\wh{\msc{L}}_{\emptyset}(s^{\prime}) \, - \,  \wh{\msc{L}}_{\mathbb{Y}}(s^{\prime})  \Big] \Bigg\} \;. 
\end{multline}

Quite similarly, one gets 
\bem
\pl{  a,  b=1 }{ N }  \f{    \sinh(\mu_a-\la_b - \i\zeta) }{ \sinh(\mu_a-\mu_b- \i\zeta)    }  \; = \;   \pl{a=1}{N} \bigg\{ \f{ \sinh(\la_a+\tfrac{\be}{N}+\i\zeta) }{ \sinh(\nu_a+\tfrac{\be}{N}+\i\zeta) } \bigg\}^N
\cdot  \pl{z ,\,  z^{\prime} \in \wh{\mathbb{Y}} }{} \bigg\{  \f{1}{   \sinh(z - z^{\prime} -\i\zeta)  } \bigg\} \\
\times \exp\Bigg\{  - \Oint{ \msc{C} }{} \dd s \,\dd s^{\prime} \,\coth^{\prime}(s-s^{\prime}-\i\zeta )\,\wh{\msc{L}}_{\mathbb{Y}}(s)\, \Big[\wh{\msc{L}}_{\emptyset}(s^{\prime}) \, - \,  \wh{\msc{L}}_{\mathbb{Y}}(s^{\prime})  \Big] \Bigg\} \\
 \times \pl{z \in \wh{\mathbb{Y}} }{}\Bigg[ \f{1}{\sinh^N( -z-\i\zeta -\tfrac{\be}{N} ) } \cdot 
 \exp\Bigg\{   \Oint{ \msc{C} }{} \dd s  \,\coth(s-z-\i\zeta) \,\wh{\msc{L}}_{\mathbb{Y}}(s)  -  \coth(s-z+\i\zeta) \,\Big[\wh{\msc{L}}_{\emptyset}(s) \, - \,   \wh{\msc{L}}_{\mathbb{Y}}(s) \Big]  \Bigg\} 
 \Bigg] \;. 
\end{multline}

Using again \eqref{ecriture produit general avec un z}, one also obtains
\begin{equation}
V_{\pm}(\om)\,=\, \pl{ z \in \wh{\mathbb{Y}} }{ } \bigg\{ \f{1}{ \sinh(z-\om\mp \i\zeta) } \bigg\} \cdot  
%
\exp\Bigg\{ -  \Oint{ \msc{C} }{} \dd s  \coth(s-\om\mp \i\zeta) \,\Big[\wh{\msc{L}}_{\emptyset}(s) \, - \,   \wh{\msc{L}}_{\mathbb{Y}}(s)  \Big]  \Bigg\}  \, 
\end{equation}
whenever $\om\pm\i\zeta\notin\msc{D}$,
\beq
\pl{a=1}{N} \f{ \sinh(\mu_a-\om)  }{  \sinh(\la_a-\om)  }  \,=\, \pl{ z \in \wh{\mathbb{Y}} }{ } \Big\{ \sinh(z-\om)   \Big\} \cdot  
 \exp\Bigg\{ -  \Oint{ \msc{C} }{} \dd s  \coth(s-\om) \,\Big[\wh{\msc{L}}_{\mathbb{Y}}(s) \, - \,   \wh{\msc{L}}_{ \emptyset}(s)  \Big]  \Bigg\}
\enq
whenever $\om   \not\in\msc{D}$, and
\beq
\pl{a=1}{N} \bigg\{ \f{ \sinh(\la_a, \i\zeta + \tfrac{\be}{N} ) }{  \sinh(\nu_a, \i\zeta + \tfrac{\be}{N} ) }  \bigg\}^{N} \; = \; 
\exp\Bigg\{   N\Oint{ \msc{C} }{} \dd s \,  \Big[\wh{\msc{L}}_{\mathbb{Y}}(s) \, - \,   \wh{\msc{L}}_{ \emptyset}(s)  \Big] \,  \Dp{s}\ln \Big\{ \sinh(s, \i\zeta + \tfrac{\be}{N} ) \Big\}  \Bigg\} \;. 
\enq

By gathering all these results, the claim follows. \qed

\subsection{Taking the infinite Trotter number limit in the complete representation}

We gather here the results of Proposition~\ref{prop-boundary-factor} and Proposition~\ref{prop-matrix-el-finite-Trotter}:

\begin{theorem}\label{th-complete-finite-Trotter}
With the hypothesis and notations of Proposition~\ref{prop-boundary-factor} and Proposition~\ref{prop-matrix-el-finite-Trotter},
%
%
%
\bem
\label{ff-finite-Trotter}
%
 \mc{F}_{\mc{B}}\big( \{ \mu_a(h^{\prime}) \}_1^N ; \{\lambda_a(h)\}_1^N; \xi_- \big) 
\cdot \f{   \Big( \bs{\Psi}( \{ \mu_a(h^{\prime}) \}_1^N ) ,   \bs{\Psi}( \{ \la_a(h) \}_1^N )  \Big)   }
{       \Big(\bs{\Psi}\big( \{ \mu_a(h^{\prime}) \}_1^N \big),  \bs{\Psi}\big( \{ \mu_a(h^{\prime}) \}_1^N \big) \Big)   } \; = \; 
%
\ex{ \wh{\msc{F}}_{\mathbb{Y}}- \wh{\msc{F}}_{\emptyset} } \cdot 
\Bigg[ \f{  1 - \wh{\mf{a}}_{\emptyset}(0)^2 }{ 1-\wh{\mf{a}}_{\mathbb{Y}}(0)^2 }\Bigg]^{\f{1}{4}}    \cdot
\Bigg[ \f{  1 +\wh{\mf{a}}_{\mathbb{Y}}(-\xi_-) }{ 1+\wh{\mf{a}}_{\emptyset}(-\xi_-) }\Bigg]^{\bs{1}_{\msc{D}}(-\xi_-) } \!\! \cdot \,
\\
%
%
%
%
\times \exp\Bigg\{ \Oint{ \msc{C} }{} \dd s  \!\! \Oint{ \msc{C}^{\prime}\subset \msc{C} }{} \!\! \dd s^{\prime}  \,
\bigg[  \phi^{(+)}_2(s,s^{\prime})\, \Big(  \wh{\msc{L}}_{\mathbb{Y}}(s)\, \wh{\msc{L}}_{\mathbb{Y}}(s^{\prime})\, -\,  \wh{\msc{L}}_{\emptyset}(s)\, \wh{\msc{L}}_{\emptyset}(s^{\prime})  \Big) 
+  \phi^{(-)}_2(s,s^{\prime})\, \wh{\msc{L}}_{\mathbb{Y}}(s)\, \Big(\wh{\msc{L}}_{\mathbb{Y}}(s^{\prime}) \, - \,  \wh{\msc{L}}_{\emptyset}(s^{\prime})  \Big) \bigg]  \Bigg\} \\
\times \exp\Bigg\{ \Oint{ \msc{C} }{} \dd s \, \phi_1(s)\, \Big( \wh{\msc{L}}_{\mathbb{Y}}(s)\, - \, \wh{\msc{L}}_{\emptyset}(s) \Big)   \Bigg\}
\cdot 
\mc{K}^{(\e{tot})}_N \big( \wh{\mathbb{Y}} \big) \cdot  \f{ (1\, - \,  \varkappa) V_+(\th)   }{ 1\, - \,\tfrac{ \wh{\mf{a}}_{\mathbb{Y}}(\th) }{ \wh{\mf{a}}_{\emptyset}(\th)  }    }  \cdot 
\f{ \det_{\Ga(\msc{C}) }\Big[ \op{id}  \; + \; \wh{\op{U}}_{\th}^{(\la)} \Big] }{   \det_{ \msc{C}_\mathbb{Y} } \Big[  \op{id}  \; + \; \wh{\ov{K}}_{\mathbb{Y}} \Big]  } \; ,
\end{multline}
in which $\mc{K}^{(\e{tot})}_N\big( \wh{\mathbb{Y}} \big)$ is expressed in terms of the particle-hole configuration
of the set of Bethe roots $\{\mu_a(h^{\prime})\}_1^{N}$ as
\begin{align}
\label{K_N^tot}
\mc{K}^{(\e{tot})}_N\big( \wh{\mathbb{Y}} \big)  \;& = \;  \mc{K}^{(1)}_N\big( \wh{\mathbb{Y}} \big)\, \mc{K}^{(2)}_N\big( \wh{\mathbb{Y}} \big) \nonumber\\
 \; & = \; \f{ \pl{a<b}{n}f(\wh{y}_a,\wh{y}_b) f(\wh{x}_a,\wh{x}_b)  }{ \pl{a,b=1}{n} f(\wh{x}_a,\wh{y}_b)   }
\cdot \f{ \pl{ \substack{z\not=z^{\prime} \\ z, z^{\prime}\in \wh{\mathbb{Y}} } }{}  \sinh(z-z^{\prime}) }{   \pl{z, z^{\prime}\in \wh{\mathbb{Y}}  }{}  \sinh(z-z^{\prime}-\i\zeta)  } 
\cdot
\pl{ x \in \wh{\mf{X}} }{} \bigg\{  \f{   \wh{\mf{a}}_{\emptyset}(x)  \, + \, 1  }{  -\ln^{\prime} \wh{\mf{a}}_{\mathbb{Y}}(x)  }   \bigg\}  \cdot 
\pl{ y \in \wh{\mc{Y}} }{} \bigg\{  \f{   1  }{  \ln^{\prime} \wh{\mf{a}}_{\mathbb{Y}}(y)  }   \bigg\}  
\nonumber\\
&\hspace{-1cm}\times 
 \pl{a=1}{n} \bigg\{  \f{ \sinh(\wh{y}_a + \xi_-) }{ \sinh(\wh{x}_a + \xi_-) } \cdot \f{ \sinh(2 \, \wh{x}_a -\i\zeta) }{ \sinh(2 \, \wh{y}_a) } \cdot
 \left[ 1+\wh{\mf{a}}_{\mathbb{Y}}(-\wh{x}_a) \right]^{\f{1}{2}}  \cdot  \left[ 1+\wh{\mf{a}}_{\mathbb{Y}}(-\wh{y}_a) \right]^{\f{1}{2}} \bigg\}  
 \nonumber\\
&\hspace{-1cm}\times 
\pl{z \in \wh{\mathbb{Y}} }{} \exp\Bigg\{ - \Oint{ \msc{C} }{}   \dd  s  \, \wh{\msc{L}}_{\mathbb{Y}}(s) \,  \Dp{s} \ln f(s,z) 
\, + \, \Oint{ \msc{C} }{} \dd s \, \wt{\psi}_1(s,z) \, \wh{\msc{L}}_{\mathbb{Y}}(s)\, + \, \wt{\psi}_2(s,z)\, \Big[\wh{\msc{L}}_{\emptyset}(s) \, - \,  \wh{\msc{L}}_{\mathbb{Y}}(s)  \Big] \Bigg\} \, .
\end{align}

\end{theorem}

It is now straightforward to take the infinite Trotter number limit of this expression, along the lines settled in Section \ref{sec-Trotter-limit}. The limiting expression
\begin{equation}\label{limit-ff}
\msc{A}^{(z)}_{h,h^{\prime}}(\mathbb{Y}) \, = \;
\lim_{N\tend +\infty} \Bigg\{  \mc{F}_{\mc{B}}\big( \{ \mu_a(h^{\prime}) \}_1^N ; \{\lambda_a(h)\}_1^N; \xi_- \big)
\cdot \f{   \Big( \bs{\Psi}( \{ \mu_a(h^{\prime}) \}_1^N ) ,   \bs{\Psi}( \{ \la_a(h) \}_1^N )  \Big)   }
{       \Big(\bs{\Psi}\big( \{ \mu_a(h^{\prime}) \}_1^N \big),  \bs{\Psi}\big( \{ \mu_a(h^{\prime}) \}_1^N \big) \Big)   } \Bigg\}   \; 
\end{equation}
is then simply given by a mere replacement $\wh{\mf{a}}_{\mathbb{Y}} \hookrightarrow \mf{a}_{\mathbb{Y}} $ and $\wh{\msc{L}}_{\mathbb{Y}} \hookrightarrow {\msc{L}}_{\mathbb{Y}}$ in
the above formulas, and by taking the limiting values of the particle and hole roots as in \eqref{limit-particle-hole-roots}.
Also (see \cite{KozPozsgaySurfaceFreeEnergyBoundaryXXZ}), the limiting expression $\msc{F}_{\mathbb{Y}}$ of $\wh{\msc{F}}_{\mathbb{Y}}$
is obtained by replacing in \eqref{ecriture msc hat F a trotter fini} the kernel $\wh{U}_{\mathbb{Y}}$ \eqref{kernelU-new} and
the function $\wh{\mf{r}}_{\mathbb{Y}}$ \eqref{fct-r-new} respectively by
\bem\label{kernelU-limit}
{U}_{\mathbb{Y}}(\om,\om^{\prime}) = \f{ - \ex{-\f{h'}{T}} \s{2\om^{\prime}-\i\zeta} }{ \s{\om+\om^{\prime}} \s{\om-\om^{\prime}-\i\zeta}}
\cdot 
\exp\big\{-2\beta\,[\coth(\om')-\coth(\om'-\i\zeta)]\big\}\\
%
%
%
\times
\exp\Bigg\{  - \Oint{ \msc{C}_U }{}  \dd  s  \, {\msc{L}}_{\mathbb{Y}}(s) \, 
\Big[ \coth(\om^{\prime} +s) \, + \, \coth( \om^{\prime} -s )
 \, -  \, \coth(\om^{\prime}+s -\i\zeta) \, - \, \coth(\om^{\prime} -s- \i\zeta)  \Big] \Bigg\}	\;, 
\end{multline}
and
\bem\label{fct-r-limit}
{\mf{r}}_{\mathbb{Y}}(\om) = \ex{ -\f{2h'}{T} } 
\cdot 
\exp\big\{-2\beta\,[\coth(\om'+\i\zeta)-\coth(\om'-\i\zeta)]\big\}\\
%
%
\times
\exp\Bigg\{  - \Oint{ \msc{C}_U }{}  \dd  s  \, {\msc{L}}_{\mathbb{Y}}(s) \, 
\Big[ \coth(s +\om +\i\zeta) \, + \, \coth(\om -s +\i\zeta)
 \, -  \, \coth(\om +s -\i\zeta) \, - \, \coth(\om -s -\i\zeta)  \Big]  \Bigg\}	\;. 
\end{multline}
The only point of attention in taking the limit in \eqref{ff-finite-Trotter} comes from the fact that, strictly speaking, the limiting value of $\wh{\mf{a}}_{\mathbb{Y}}(0)$ is not well defined. However, the limiting value of its square $\wh{\mf{a}}_{\mathbb{Y}}(0)^2$ can be taken without problem from the integral equation
 \label{ecriture eqn NLI forme primordiale}
\beq
\wh{\mf{a}}_{\mathbb{Y}}(0)^2 \; = \;   \bigg(  \f{  \sinh(   \tf{ \be }{N} -\i\zeta) }{    \sinh( - \tf{ \be }{N}  - \i\zeta )  }   \bigg)^{2N} \ex{-2\f{h'}{T}}
  \pl{ y \in \wh{\mathbb{Y}} }{} \Big\{ \ex{2\i  \th(-y) } \Big\} \cdot  \exp\bigg\{  4\i\pi \Oint{  \msc{C}  }{} K( u)\, \wh{\msc{L}}_{\mathbb{Y}}(u)\, 
  \dd u   \bigg\} \; ,
\enq
from which it follows that
\beq
\label{limit-a^2(0)}
\lim_{N \tend + \infty} \Big\{ \,\wh{\mf{a}}_{\mathbb{Y}}(0)^2 \Big\} \; =  \; \mc{A}\big(\mathbb{Y} \big)
\enq
is well defined and finite.

\section{The thermal form factor expansion for the one point function}
\label{sec-result}

From the previous results, one therefore obtains the below thermal form factor
series expansion for the generating function $\mc{Q}(h^{\prime},m) \, = \, \underset{N\tend + \infty}{\lim} \mc{Q}_N(h^{\prime},m)$ of the one-point
function at distance $m$ from the boundary. One has 
\beq
\label{one-point-fct-result}
\big< \sg_{m+1}^{z} \big>_{T} \; = \;      2 T  \Dp{h^{\prime}} \mf{D}_m \mc{Q}(h^{\prime},m)_{\mid h^{\prime}=h}
\enq
where 
\beq
\mc{Q}(h^{\prime},m) \, = \, \sul{\mathbb{Y} }{}    \bigg( \f{ \tau_{\mathbb{Y}}(0) }{  \tau_{\emptyset}(0)  } \bigg)^{m} \msc{A}^{(z)}_{h,h^{\prime}}(\mathbb{Y}) \;. 
\enq
Above, the summation runs through all possible particle-hole excitations, \textit{viz}. though all the solutions to the non-linear integral equation 
\eqref{ecriture NLIE Trotter infini}
subject to the conditions which fix a given particle-hole excitations \eqref{ecriture condition subsidiaires solution NLIE spectre}, and 
$\msc{A}^{(z)}_{h,h^{\prime}}(\mathbb{Y})$
denotes the infinite Trotter limit \eqref{limit-ff} of the product \eqref{ff-finite-Trotter}. 
Defining $\msc{C}=\mc{C} + \i\tfrac{\zeta}{2}$,  $ \mc{D} = \e{Int}\big( \mc{C} \big)$, and setting
\beq
\mc{L}_{\mathbb{Y}}(\la) \; = \; \f{1}{2\i\pi} \msc{L}n\Big[ 1+ \ex{-\f{1}{T}u_{\mathbb{Y}}(\la)  }\Big] \;, \quad \e{with} \quad 
u_{\mathbb{Y}}(\la) \, = \, - T \mf{A}_{\mathbb{Y}}(u+\i\tfrac{\zeta}{2}) \; ,
\enq
%
%
%
this quantity $\msc{A}^{(z)}_{h,h^{\prime}}(\mathbb{Y})$ can explicitly be rewritten as
\begin{equation}\label{ff-infinite-Trotter}
 \msc{A}^{(z)}_{h,h^{\prime}}(\mathbb{Y}) \, 
 = \, 
\ex{ \msc{F}_{\mathbb{Y}} \, - \,  \msc{F}_{\emptyset} }  \cdot \Big( \mc{E}^{(r)}\cdot \mc{K}^{(r)}\Big)\big( \mathbb{Y} \big) \cdot \Big( \mc{E}^{(s)}\cdot \mc{K}^{(s)}\Big)\big( \mathbb{Y} \big)\, .
\end{equation}
Here we have introduced
\bem
 \mc{E}^{(r)}\big( \mathbb{Y} \big) \; = \; \exp\Bigg\{ -\Oint{ \mc{C} }{} \dd \la  \Oint{ \mc{C}^{\prime}\subset \mc{C} }{} \dd \mu 
\bigg[ \f{1}{2} \coth^{\prime}(\la + \mu + \i\zeta) \cdot  \Big(  \mc{L}_{\mathbb{Y}}(\la)\, \mc{L}_{\mathbb{Y}}(\mu)\, -\,  \mc{L}_{\emptyset}(\la)\,\mc{L}_{\emptyset}(\mu)  \Big) \\
+   \coth^{\prime}(\la - \mu - \i \zeta) \cdot  \mc{L}_{\mathbb{Y}}( \la )\, \Big(\mc{L}_{\emptyset}( \mu ) \, - \,  \mc{L}_{\mathbb{Y}}( \mu )  \Big) \bigg]  \Bigg\} 
\cdot \exp\Bigg\{ \lim_{\eps\tend 0^+} \Oint{ \mc{C} }{} \dd \la \wt{\phi}_1(\la+\i\eps)\Big( \mc{L}_{\mathbb{Y}}(\la)\, - \,\mc{L}_{\emptyset}(\la) \Big)   \Bigg\}\, ,
\label{definition E regulier}
\end{multline}
in which we agree upon 
\beq
\wt{\phi}_1(\la) \, = \, \coth(2\la) \, + \, \coth\big(2\la+\i\zeta\big) \, - \,\coth\big( \la + \xi_- + \i\tfrac{\zeta}{2} \big) \; ,
\enq
and
\bem
 \mc{E}^{(s)}\big( \mathbb{Y} \big) \; = \; \exp\Bigg\{ \Oint{ \mc{C} }{} \dd \la  \Oint{ \mc{C}^{\prime}\subset \mc{C} }{} \dd \mu
\bigg[ \f{1}{2} \coth^{\prime}(\la+\mu) \cdot  \Big(  \mc{L}_{\mathbb{Y}}(\la)\, \mc{L}_{\mathbb{Y}}(\mu)\, -\,  \mc{L}_{\emptyset}(\la)\,\mc{L}_{\emptyset}(\mu)  \Big) \\
+   \coth^{\prime}(\la-\mu) \cdot  \mc{L}_{\mathbb{Y}}( \la )\, \Big(\mc{L}_{\emptyset}( \mu ) \, - \,  \mc{L}_{\mathbb{Y}}( \mu )  \Big) \bigg]  \Bigg\} \;. 
\end{multline}
We have also defined
\bem
 \mc{K}^{(r)}\big( \mathbb{Y} \big) \; = \;  \bigg( \f{ 1-  \mc{A}\big(\emptyset \big) }{  1 -   \mc{A}\big(\mathbb{Y} \big)  }\bigg)^{\f{1}{4}} \cdot   \f{ \det_{\Ga(\msc{C}) }\Big[ \op{id}  \; + \;  \op{U}_{\th}^{(\la)} \Big] }{   \det_{ \msc{C}_\mathbb{Y} } \Big[  \op{id}  \; + \; \ov{K}_{\mathbb{Y}} \Big]  } 
\cdot \f{ (1\, - \,  \varkappa) V_+(\th)   }{ 1\, - \,\tfrac{ \mf{a}_{\mathbb{Y}}(\th) }{ \mf{a}_{\emptyset}(\th)  }    } 
\cdot  \pl{ a=1 }{ n } \bigg\{  \f{ - 1 \, - \,  \ex{- \f{1}{T}u_{\emptyset}(x_a)  }  }{   \wh{u}^{\,\prime}_{\mathbb{Y}}(x_a) \cdot \wh{u}^{\,\prime}_{\mathbb{Y}}(y_a)  }   \bigg\} 
\cdot \pl{z \in \mc{Y}\ominus \mf{X} }{}  \Big\{ \mc{B}^{(r)}(z) \Big\} \\
\times   \pl{a=1}{n} \bigg\{  \f{ \sinh(y_a + \xi_- + \i \tf{\zeta}{2}) }{ \sinh(x_a + \xi_- + \i \tf{\zeta}{2}) } \cdot \f{ \sinh(2 x_a ) }{ \sinh(2 \, y_a+ \i \zeta) } 
 \cdot \Big\{ 1+   \ex{- \f{1}{T}u_{\mathbb{Y}}(-x_a-\i\zeta)  }  \Big\}^{\f{1}{2}}  \cdot  \Big\{ 1+   \ex{- \f{1}{T}u_{\mathbb{Y}}(-y_a-\i\zeta)  }  \Big\}^{\f{1}{2}} \bigg\}  \\
\times \pl{a,b=1}{n} \bigg\{ \f{ \sinh(x_a+y_b+\i\zeta) \sinh(x_a-y_b,\i\zeta) }{  \sinh(x_a-x_b-\i\zeta) \sinh(y_a-y_b-\i\zeta)   }   \bigg\}
\cdot 
\pl{a<b}{n} \bigg\{ \f{ 1 }{  \sinh(x_a+x_b+\i\zeta) \sinh(y_a+y_b+\i\zeta)   }   \bigg\}  \\
 \times \Bigg(  \f{  1  +   \ex{- \f{1}{T}u_{\mathbb{Y}}(-\xi_- - \i\tf{\zeta}{2})  } }{  1  +   \ex{- \f{1}{T}u_{\emptyset}(-\xi_- - \i\tf{\zeta}{2})  } }  \Bigg)^{ \bs{1}_{ \mc{D} }( -\xi_- - \i\tf{\zeta}{2})  }  \; .
\label{definition K regulier}
\end{multline}
%
%
%
Here the integral kernels $\op{U}_{\th}^{(\la)} $ and $\ov{K}_{\mathbb{Y}}$ are obtained upon the replacement $\wh{\mf{a}}_{\mathbb{Y}} \hookrightarrow \mf{a}_{\mathbb{Y}} $
in the associated expressions at finite Trotter number \eqref{definition noyau integral hat U th lambda} and \eqref{definition noyau integral hat ov K Y}, the quantity $\mc{A}\big(\mathbb{Y} \big)$ is given by (see \eqref{limit-a^2(0)})
\beq
  \mc{A}\big(\mathbb{Y} \big) \, = \, 
  \pl{ y\in \mc{Y}\ominus \mf{X}  }{} \Big\{ \ex{ 2 \i  \th(-y-\i\f{\zeta}{2}) } \Big\} \cdot  \exp\bigg\{ -2 \f{h}{T} + 4 \be \coth(\i\zeta) \, + \,  4\i\pi \Oint{  \mc{C}  }{} K(\la+\i\tfrac{\zeta}{2}) \cdot  \mc{L}_{\mathbb{Y}}(\la)  \cdot \dd \la   \bigg\}  \, ,
\enq
and we have introduced 
\beq
\mc{B}^{(r)}(z) \, = \,  \exp\Bigg\{  \Oint{ \mc{C} }{}   \dd  \la \bigg[  \mc{L}_{\mathbb{Y}}(\la) \Big[  \coth(\la-z-\i\zeta) + \coth(\la+z+\i\zeta) \Big]  \, - \,
\coth(\la-z+\i\zeta) \Big(\mc{L}_{\emptyset}(\la) \, - \,  \mc{L}_{\mathbb{Y}}(\la)  \Big) \bigg] \Bigg\} \; .
\enq
Further, one has
\begin{equation}
V_+(\th)\,=\, \pl{ z  \in \mc{Y}\ominus \mf{X}  }{ } \bigg\{ \f{1}{ \sinh(z-\th- \tf{ \i\zeta }{ 2} ) } \bigg\} \cdot
%
\exp\Bigg\{ -  \Oint{ \mc{C} }{} \dd s \, \coth\Big( s - \th - \f{\i\zeta}{2} \Big) \,\Big[\wh{\msc{L}}_{\emptyset}(s) \, - \,   \wh{\msc{L}}_{\mathbb{Y}}(s)  \Big]  \Bigg\}  \, .
\end{equation}

Finally, the last factor appearing in \eqref{ff-infinite-Trotter} takes the form 
\beq
 \mc{K}^{(s)}\big( \mathbb{Y} \big) \; = \; T^{2n} \cdot \pl{z \in \mc{Y}\ominus \mf{X}  }{}  \Big\{ \mc{B}^{(s)}(z) \Big\} \cdot
\f{ \pl{a<b}{n} \Big\{ \sinh^2(x_a-x_b) \sinh^2(y_a-y_b)     \sinh(x_a+x_b) \sinh(y_a+y_b)      \Big\}   }{ \pl{a,b=1}{n}  \Big\{ \sinh^2(x_a-y_b) \sinh(x_a+y_b) \Big\}  } \; ,
\enq
which involves the function
\beq
\mc{B}^{(s)}(z) \, = \,  \exp\Bigg\{  \Oint{ \mc{C} }{}   \dd  \la \bigg[ \coth(\la-z) \Big(\mc{L}_{\emptyset}(\la) \, - \,  \mc{L}_{\mathbb{Y}}(\la)  \Big)  
\, - \, \mc{L}_{\mathbb{Y}}(\la) \Big[ \coth(\la+z) + \coth(\la-z) \Big]  \bigg] \Bigg\}  \;.
\enq

We would also like to recall that this representation holds when the parameter $\th$ is such that $\th + \i\tf{\zeta}{2} \not \in \e{Int}(\mc{C})$ and that
$\varkappa = \ex{ \f{ h - h^{\prime} }{ T } }$.


\section{Conclusion}

This work develops a setting allowing one to produce thermal form factor expansions for multi-point correlation functions in open quantum integrable models. 
Such kinds of thermal form factor expansions, which had been so far only developed in the periodic case, have proven to be very efficient for the computation of correlation functions, and notably for the determination of their long-distance asymptotic behaviour.  It is worth recalling that, for integrable models with open boundary conditions, the only exact representations that could have been obtained so far for the correlation functions turned out to be too intricate for allowing any asymptotic analysis. We therefore hope that the new approach that we propose here will open the way to solve this long-standing problem.

Our method relies on the ideas developed in \cite{KozDugaveGohmannThermaxFormFactorsXXZ,KozPozsgaySurfaceFreeEnergyBoundaryXXZ}. We have here explicitly worked out the
expansion describing the one-point function at distance $m$ from the boundary in the XXZ spin-1/2 chain with diagonal boundary fields.  It is however clear that the thermal form factor expansion for multi-point correlation functions can be worked out completely similarly. We also stress that, by conforming the techniques of \cite{KozGohmannKarbachSuzukiFiniteTDynamicalCorrFcts}, it is possible to generalise our setting to dynamical correlation functions as well. Let us finally mention that our method is {\em a priori} applicable to all 
quantum integrable models whose thermodynamics can be addressed within the quantum transfer matrix method. 

The next problem that we would like to consider is the analysis of the low-temperature limit of our final result.
Indeed, similarly as what happens in the periodic case \cite{KozDugaveGohmannThermaxFormFactorsXXZ,KozDugaveGohmannThermaxFormFactorsXXZOffTransverseFunctions},
we expect to be able to grasp, from this form factor expansion,
the asymptotic behaviour of the one-point functions \eqref{one-point-fct-result} in terms of the distance $m$ from the boundary.
To this aim, we would like to mention that our final result has already been explicitly decomposed into two parts: one which has a finite limit when $T\tend 0^+$,
the factors $\mc{E}^{(r)}(\mathbb{Y})$, $\mc{K}^{(r)}(\mathbb{Y})$,
and one which we expect to produce a power-law behaviour in $T$, the factors $\mc{E}^{(s)}(\mathbb{Y})$, $\mc{K}^{(s)}(\mathbb{Y})$. The latter should be responsible for the emergence of a conformal field theoretic description of the long-distance asymptotic behaviour of the one point function. We plan to consider this interesting problem in a next publication.

\section*{Acknowledgment}

K.K.K. and V.T. acknowledge support from  CNRS and are indebted to F. Göhmann for stimulating discussions.


\begin{thebibliography}{10}

\bibitem{BogoliubiovIzerginKorepinBookCorrFctAndABA}
N.M. Bogoliubov, A.G. Izergin, and V.E. Korepin, \emph{{"Quantum inverse
  scattering method, correlation functions and algebraic Bethe Ansatz."}},
  Cambridge monographs on mathematical physics, 1993.

\bibitem{BortzFrahmGohmannSurfaceFreeEnergy}
M.~Bortz, H.~Frahm, and F.~G\"{o}hmann, \emph{{"Surface free energy for systems
  with integrable boundary conditions."}}, J. Phys. A: Math. Gen. \textbf{\bf
  38} (2005), 10879--10892.

\bibitem{CherednikReflectionEquationFactorisabilityOfScattering}
V.I. Cheredink, \emph{{"Factorizing particles on a half-line and root
  systems."}}, Theor. Math. Phys. \textbf{\bf{61}} (1984), 977.

\bibitem{DestriDeVegaAsymptoticAnalysisCountingFunctionAndFiniteSizeCorrectionsinTBAFirstpaper}
C.~Destri and H.J. de~Vega, \emph{{"New thermodynamic Bethe Ansatz equations
  without strings."}}, Phys. Rev. Lett. \textbf{\bf{69}} (1992), 2313--2317.

\bibitem{KozDugaveGohmannThermaxFormFactorsXXZ}
M.~Dugave, F.~G\"{o}hmann, and K.K. Kozlowski, \emph{{"Thermal form factors of
  the XXZ chain and the large-distance asymptotics of its temperature dependent
  correlation functions."}}, J. Stat. Mech. \textbf{1307} (2013), P07010.

\bibitem{KozDugaveGohmannThermaxFormFactorsXXZOffTransverseFunctions}
\bysame, \emph{{"Low-temperature large-distance asymptotics of the transversal
  two-point functions of the XXZ chain."}}, J. Stat. Mech. \textbf{1404}
  (2014), P04012.

\bibitem{FaddeevSklyaninTakhtajanSineGordonFieldModel}
L.D. Faddeev, E.K. Sklyanin, and L.A. Takhtadzhan, \emph{{"Quantum inverse
  problem method I."}}, Teor. Math. Phys. \textbf{\bf{40}} (1979), 688--706.

\bibitem{KozGohmannKarbachSuzukiFiniteTDynamicalCorrFcts}
F.~G\"{o}hmann, M.~Karbach, A.~Kl\"{u}mper, K.K. Kozlowski, and J.~Suzuki,
  \emph{{"Thermal form-factor approach to dynamical correlation functions of
  integrable lattice models."}}, J. Stat. Mech. (2017), 113106.

\bibitem{GohmannKlumperSeelFinieTemperatureCorrelationFunctionsXXZ}
F.~G\"{o}hmann, A.~Kl\"{u}mper, and A.~Seel, \emph{{"Integral representations
  for correlation functions of the XXZ chain at finite temperature."}}, J.
  Phys. A: Math. Gen. \textbf{\bf 37} (2004), 7625--7652.

\bibitem{GrijalvaDeNArdisTerrasXXZBoundaryMagnetisation}
S.~Grijalva, J.~De Nardis, and V.~Terras, \emph{{"Open XXZ chain and boundary
  modes at zero temperature."}}, SciPost Phys. \textbf{\bf 7} (2019), 023,
  77pp.

\bibitem{KozGohmannGoomaneeSuzukiRigorousApproachQTMForFreeEnergy}
F.~Göhmann, S.~Goomanee, K.K. Kozlowski, and J.~Suzuki, \emph{{"Thermodynamics
  of the spin-1/2 Heisenberg-Ising chain at high temperatures: a rigorous
  approach."}}, Comm. Math. Phys. \textbf{377} (2020), 623--673.

\bibitem{HultenGSandEnergyForXXX}
L.~H\'{u}lthen, \emph{{"\"{U}ber das Austauschproblem eines Kristalles."}},
  Arkiv Mat. Astron. Fys. \textbf{\bf{26A}} (1938), 1--106.

\bibitem{IzerginKorepinQISMApproachToCorrFns2SiteModel}
A.G. Izergin and V.E. Korepin, \emph{{"The quantum inverse scattering method
  approach to correlation functions."}}, Comm. Math. Phys. \textbf{\bf{94}}
  (1984), 67--92.

\bibitem{IzerginKorepinQISMApproachToCorrNextDiscussion}
\bysame, \emph{{"Correlation functions for the Heisenberg XXZ
  antiferromagnet."}}, Comm. Math. Phys. \textbf{\bf{99}} (1985), 271--302.

\bibitem{JimboKedemKonnoMiwaXXZChainWithaBoundaryElemBlcks}
M.~Jimbo, R.~Kedem, T.~Kojima, H.~Konno, and T.~Miwa, \emph{{"XXZ chain with a
  boundary."}}, Nucl. Phys. B \textbf{\bf 441} (1995), 437--470.

\bibitem{JimboMikiMiwaNakayashikiElementaryBlocksXXZperiodicDelta>1}
M.~Jimbo, K.~Miki, T.~Miwa, and A.~Nakayashiki, \emph{{"Correlation functions
  of the XXZ model for $\Delta<-1$."}}, Phys. Lett. A \textbf{\bf 168} (1992),
  256--263.

\bibitem{JimboMiwaFormFactorsInMassiveXXZ}
M.~Jimbo and T.~Miwa, \emph{{"Algebraic analysis of solvable lattice models"}},
  Conference Board of the Mathematical Sciences, American Mathematical Society,
  1995.

\bibitem{JimboMiwaElementaryBlocksXXZperiodicMassless}
\bysame, \emph{{"QKZ equation with $\mid q \mid$ =1 and correlation functions
  of the XXZ model in the gapless regime."}}, J. Phys. A \textbf{\bf 29}
  (1996), 2923--2958.

\bibitem{KozKitMailNicSlaTerElemntaryblocksopenXXZ}
N.~Kitanine, K.K. Kozlowski, J.-M. Maillet, G.~Niccoli, N.A. Slavnov, and
  V.~Terras, \emph{{"Correlation functions of the open XXZ chain I."}}, J.
  Stat. Mech.: Th. and Exp. (2007), P10009.

\bibitem{KozKitMAilNicSlaTerResummationsOpenXXZ}
\bysame, \emph{{"Correlation functions of the open XXZ chain II."}}, J. Stat.
  Mech.: Th. and Exp. (2008), P07010.

\bibitem{KozKitMailSlaTer6VertexRMatrixMasterEquation}
N.~Kitanine, K.K. Kozlowski, J.-M. Maillet, N.A. Slavnov, and V.~Terras,
  \emph{{"On correlation functions of integrable models associated with the
  six-vertex R-matrix."}}, J. Stat. Mech. (2007), P01022.

\bibitem{KozKitMailSlaTerXXZsgZsgZAsymptotics}
\bysame, \emph{{"Algebraic Bethe Ansatz approach to the asymptotics behavior of
  correlation functions."}}, J. Stat. Mech: Th. and Exp. \textbf{04} (2009),
  P04003.

\bibitem{KozKitMailSlaTerEffectiveFormFactorsForXXZ}
\bysame, \emph{{"On the thermodynamic limit of form factors in the massless XXZ
  Heisenberg chain."}}, J. Math. Phys. \textbf{50} (2009), 095209.

\bibitem{KozKitMailSlaTerRestrictedSums}
\bysame, \emph{{"A form factor approach to the asymptotic behavior of
  correlation functions in critical models."}}, J. Stat. Mech. : Th. and Exp.
  \textbf{1112} (2011), P12010.

\bibitem{KozKitMailSlaTerThermoLimPartHoleFormFactorsForXXZ}
\bysame, \emph{{"Thermodynamic limit of particle-hole form factors in the
  massless XXZ Heisenberg chain."}}, J. Stat. Mech. : Th. and Exp.
  \textbf{1105} (2011), P05028.

\bibitem{KozKitMailSlaTerRestrictedSumsEdgeAndLongTime}
\bysame, \emph{{"Form factor approach to dynamical correlation functions in
  critical models."}}, J. Stat. Mech. \textbf{1209} (2012), P09001.

\bibitem{KitanineMailletSlavnovTerrasOriginalSeries}
N.~Kitanine, J.-M. Maillet, N.A. Slavnov, and V.~Terras, \emph{{"Spin-spin
  correlation functions of the XXZ-$1/2$ Heisenberg chain in a magnetic
  field."}}, Nucl. Phys. B \textbf{641} (2002), 487--518.

\bibitem{KitanineMailletSlavnovTerrasDynamicalCorrelationFunctions}
\bysame, \emph{{"Dynamical correlation functions of the XXZ spin-$1/2$
  chain."}}, Nucl. Phys. B \textbf{\bf 729} (2005), 558--580.

\bibitem{KitanineMailletSlavnovTerrasMasterEquation}
\bysame, \emph{{"Master equation for spin-spin correlation functions of the XXZ
  chain."}}, Nucl. Phys. B \textbf{\bf 712} (2005), 600--622.

\bibitem{KitanineMailletTerrasFormfactorsperiodicXXZ}
N.~Kitanine, J.-M. Maillet, and V.~Terras, \emph{{"Form factors of the XXZ
  Heisenberg spin-$1/2$ finite chain."}}, Nucl. Phys. B \textbf{554} (1999),
  647--678.

\bibitem{KitanineMailletTerrasElementaryBlocksPeriodicXXZ}
\bysame, \emph{{"Correlation functions of the XXZ Heisenberg spin-$1/2$ chain
  in a magnetic field."}}, Nucl. Phys. B \textbf{567} (2000), 554--582.

\bibitem{KlumperNLIEfromQTMDescrThermoXYZOneUnknownFcton}
A.~Kl\"{u}mper, \emph{{"Thermodynamics of the anisotropic spin-$1/2$ Heisenberg
  chain and related quantum chains."}}, Z. Phys. B: Cond. Mat. \textbf{\bf 91}
  (1993), 507--519.

\bibitem{KorepinNormBetheStates6-Vertex}
V.E. Korepin, \emph{{"Calculation of norms of Bethe wave-functions."}}, Comm.
  Math.Phys. \textbf{\bf 86} (1982), 391--418.

\bibitem{KozMasslessFFSeriesXXZ}
K.K. Kozlowski, \emph{{"On the thermodynamic limit of form factor expansions of
  dynamical correlation functions in the massless regime of the XXZ spin $1/2$
  chain."}}, J. Math. Phys. "Ludwig Faddeev memorial volume" \textbf{59}
  (2018), 091408.

\bibitem{KozLongDistanceLargeTimeXXZ}
\bysame, \emph{{"Long-distance and large-time asymptotic behaviour of dynamic
  correlation functions in the massless regime of the XXZ spin-1/2 chain."}},
  J. Math. Phys. \textbf{60} (2019), 073303.

\bibitem{KozSingularitiesSpectralFctsXXZ}
\bysame, \emph{{"On singularities of dynamic response functions in the massless
  regime of the XXZ spin-1/2 chain."}}, J. Math. Phys \textbf{62} (2021),
  063507, 93 pp.




\bibitem{KozFaulmanGohmannLowTNLIERigourousAnalysisForQTMMasslessRegime}
S.~Faulmann, F.~G\"{o}hmann and K.K.~Kozlowski,
    \emph{{"Low-temperature spectrum of the quantm transfer matrix of the XXZ chain in the massless regime."}}, to appear.



\bibitem{KozPozsgaySurfaceFreeEnergyBoundaryXXZ}
K.K. Kozlowski and B.~Pozsgay, \emph{{"Surface free energy of the open XXZ
  spin-1/2 chain."}}, J. Stat. Mech. {\bf 2012}, (2012), P05021.


\bibitem{PozsgayRakosBoundaryFreeEnergyGeneralBCXXZChain}
B.~Pozsgay and O.~Rákos, \emph{{"Exact boundary free energy of the open XXZ chain with arbitrary boundary conditions."}},
J. Stat. Mech. {\bf 2018}, (2018), 113102.

\bibitem{SklyaninABAopenmodels}
E.K. Sklyanin, \emph{{"Boundary conditions for integrable quantum systems."}},
  J. Phys. A: Math. Gen. \textbf{\bf 28} (1988), 2375--2389.

\bibitem{SlavnovScalarProductsXXZ}
N.A. Slavnov, \emph{{"Calculation of scalar products of wave-functions and
  form-factors in the framework of the algebraic Bethe Ansatz."}}, Theor. Math.
  Phys. \textbf{\bf 79} (1989), 502--508.

\bibitem{TakahashiSpinSpinSecondNeighborXXX}
M.~Takahashi, \emph{"half-filed hubbard model at low temperature."}, J. Phys. C
  \textbf{\bf 10} (1977), 1289--1301.

\bibitem{TsuchiyaPartitionFunctWithReflecEnd}
O.~Tsuchiya, \emph{{"Determinant formula for the six-vertex model with
  reflecting end."}}, J. Phys. A: Math. Gen. \textbf{\bf 39} (1998),
  5946--5951.

\end{thebibliography}
\end{document}